\documentclass[preprint,12pt]{elsarticle}

\usepackage{graphicx,pstricks}
\usepackage{graphics}
\usepackage{amssymb}
\usepackage{amsmath}
\usepackage{bm}
\usepackage{upgreek}
\usepackage{amsthm}
\usepackage{mathrsfs}
\usepackage{rotating}
\usepackage{tabularx} % in preamble
\usepackage{amssymb,amsfonts}
\usepackage{CJKutf8}
\usepackage[colorlinks=true,
            linkcolor=blue,
            citecolor=blue,
            urlcolor=blue]{hyperref}

\usepackage[capitalize,nameinlink,noabbrev]{cleveref}

\journal{Ultramicroscopy}

\begin{document}

\begin{frontmatter}



\title{Information in 4D-STEM: Where it is, and How to Use it} %% Article title

\author{Desheng Ma\corref{cor1}}
\ead{dm852@cornell.edu}
\affiliation{organization={School of Applied and Engineering Physics, Cornell university},%Department and Organization
            city={Ithaca},
            postcode={14853}, 
            state={NY},
            country={US}
            }

\author{Guanxing Li}
% \ead{dm852@cornell.edu}
\affiliation{organization={School of Applied and Engineering Physics, Cornell university},%Department and Organization
            city={Ithaca},
            postcode={14853}, 
            state={NY},
            country={US}
            }

\author{David A. Muller} %% Author name
\affiliation{organization={School of Applied and Engineering Physics, Cornell university},%Department and Organization
            city={Ithaca},
            postcode={14853}, 
            state={NY},
            country={US}
            }
\affiliation{organization={Kavli Institute at Cornell for Nanoscale Science},%Department and Organization
            city={Ithaca},
            postcode={14853}, 
            state={NY},
            country={US}
            }

\author{Steven E. Zeltmann\corref{cor2}} %% Author name
% \cortext[cor2]{}
\ead{steven.zeltmann@cornell.edu}
\affiliation{organization={School of Applied and Engineering Physics, Cornell university},%Department and Organization
            city={Ithaca},
            postcode={14853}, 
            state={NY},
            country={US}
            }

\affiliation{organization={Platform for the Accelerated Realization, Analysis, and Discovery of Interface Materials},%Department and Organization
            city={Ithaca},
            postcode={14850}, 
            state={NY},
            country={US}
            }
%% Abstract
\begin{abstract}
Contrast transfer mechanisms for electron scattering have been extensively studied in transmission electron microscopy. Here we revisit H. Rose’s generalized contrast formalism from scattering theory to understand where information is encoded in four-dimensional scanning transmission electron microscopy (4D-STEM) data, and consequently identify new imaging modes that can also serve as crude but fast approximations to ptychography.  We show that tilt correction and summation of the symmetric and antisymmetric scattering components within the bright-field disk—corresponding to tilt-corrected bright field (tcBF) and tilt-corrected differential phase contrast (tcDPC) respectively—enables aberration-corrected, bright-field phase contrast imaging (acBF) that makes maximal use of the 4D-STEM information under the weak phase object approximation (WPOA). Beyond the WPOA, we identify the contrast transfer from the interference between inelastic/plural scattering electrons, which show up as quadratic terms, and show that under overfocus conditions, contrast can be further enhanced at selected frequencies, similar to phase-contrast TEM imaging. There is also usable information encoded in the dark field region which we demonstrate by constructing a tilt-corrected dark-field image (tcDF) that sums up the incoherent scattering components and holds promise for depth sectioning of strong scatterers. This framework generalizes phase contrast theory in conventional/scanning transmission electron microscopy to 4D-STEM and provides analytical models and insights into full-field iterative ptychography, which blindly exploits all above contrast mechanisms.

\end{abstract}


% \begin{highlights}
% \end{highlights}

%% Keywords
\begin{keyword}



\end{keyword}

\end{frontmatter}


% \chapter{}
\section{Introduction}
Aberration-corrected electron microscopy has emerged as an essential tool for high-resolution imaging of materials, enabling direct visualization of structural, electronic, and magnetic properties at the atomic scale. In scanning transmission electron microscopy (STEM), a wide variety of imaging modes, such as bright-field (BF) \cite{zernikePhaseContrastNew1942}, high-angle annular dark-field (HAADF)) \cite{crewe1970visibility}, annular bright-field (ABF) \cite{findlay2010dynamics}, and differential phase contrast (DPC)) \cite{dekkers_differential_1974}, can be accessed by selecting different detector geometries and acquisition conditions \cite{rose_nonstandard_1976, rose1974phase, kirkland_advanced_2020}. These modes have traditionally provided complementary contrast mechanisms, from mass-thickness to electrostatic and magnetic phase gradients.

The advent of four-dimensional STEM (4D-STEM)), which records the full convergent beam electron diffraction pattern at each scan position by a pixelated array detector \cite{tate2016high, philipp2022very}, has opened new possibilities for imaging and analysis. A review of early work can be found in \cite{ophus2019four}. By capturing the full momentum-space scattering information, 4D-STEM enables virtual imaging under any conventional integrative detector geometry, but also unlocks a range of computational phase retrieval methods in electron microscopy including direct ptychography \cite{hoppe_beugung_1969,hoppe_three-dimensional_1980}, e.g., single side band (SSB) \cite{pennycook_efficient_2015}, Wigner distribution deconvolution (WDD) \cite{rodenburg1992theory, bates_sub-angstrom_1989}, and iterative ptychography \cite{maiden2009improved,maiden2017further,thibault_high-resolution_2008}, which recovers the probe and object simultaneously by optimizing a loss function defined as the difference between the forward diffraction patterns and the measured experimental data and can exceed the diffraction limit.

The direct ptychography methods including SSB and WDD were developed considering a phase-object convolution model for the electron scattering process and retrieve the phase of the sample transmission function from measured diffraction intensities using a non-iterative inversion method \cite{yang2016enhanced,yang2017electron}. This approach yields a computationally straightforward reconstruction procedure which is often able to be evaluated live as data streams from the detector \cite{bangun2023wigner, strauch2021live}. These direct ptychography methods make use of the interference between the diffracted waves and the direct wave, and thus favor a focused-probe condition to achieve optimal contrast transfer efficiency \cite{o2021contrast}. An analytical phase contrast transfer function (CTF) can be derived under the weak phase object approximation (WPOA), e.g., for SSB \cite{pennycook_efficient_2015}. Many thin specimens, particularly thin biological samples and light-element materials can be modeled as weak phase objects \cite{cowley1979adjustment}. An analysis of the information transfer, including propagation of the detector noise and comparison with Zernike phase plate imaging in conventional TEM, is presented by Bennemann et al \cite{bennemann_detective_2025}. However, these methods’ dependence on the phase object approximation and their neglect of change in the shape of the beam in the sample leads to artifacts for heavy elements and thicker samples \cite{hofer2024phase,gao2024central}.
 
Iterative ptychography methods including ePIE \cite{maiden2009improved} and multislice electron ptychography \cite{chen_electron_2021, odstrvcil2018iterative, maiden_ptychographic_2012} provide the means to handle thicker, more-strongly scattering samples \cite{chen_electron_2021}, recover arbitrarily aberrated or incoherent probes \cite{chen2020mixed}, and reconstruct information beyond the optical diffraction limit of the instrument \cite{jiang_electron_2018,chen_imaging_2024}. Iterative ptychography methods can benefit from the use of a defocused probe especially at low dose, as observed in simulations \cite{jiang_electron_2018} and later realized experimentally \cite{chen2020mixed}. The contrast transfer behavior of these methods, however, is not known analytically, with most reported results being derived from reconstructions of simulated single atoms \cite{o2021contrast, ma_information_2024}, random configurations of atoms \cite{ma_information_2024} or white noise \cite{zhou2020low}. Pseudo-CTFs derived by Terzoudis-Lumsden et al \cite{terzoudis-lumsden_resolution_2023} establish the limits of information transfer within the WPOA in three dimensions and highlight the dependence on the convergence angle of the probe as well as the collection angle of the pixelated detector. Theoretical work by Dwyer and Paganin has also attempted to establish upper bounds on the information content of a 4D-STEM dataset under the WPOA \cite{dwyer_quantum_2024}.
 
Another simple approach to recovering coherent phase contrast information from defocused 4D-STEM is tilt-corrected bright-field STEM imaging (tcBF-STEM) \cite{nguyen_4d-stem_2016,spoth_dose-efficient_2017,yu_dose-efficient_2024}, which operates by correcting the relative shifts between brightfield images produced from each detector pixel within the bright-field disk which are a consequence of the applied defocus \cite{spoth_dose-efficient_2017,yu_dose-efficient_2024}. It is sometimes also referred to as parallax imaging \cite{varnavides_iterative_2023, kucukoglu_low-dose_2024}, although this can be confusing as a tilt-corrected DPC-based approach with the same name also exists \cite{seifer_flexible_2021}, or shadow montage imaging \cite{seiferShadowMontageConeBeam2025}. TcBF-STEM works well at very low dose (< 1e$^-$/\AA$^2$) and through very thick (> 500 nm) non-crystalline samples, and via upsampling can provide reconstructions with spatial sampling finer than the raster scan of the probe \cite{yu_dose-efficient_2024}. Yu et al \cite{yu_dose-efficient_2024} also analyzed the contrast transfer of this method in the WPOA and found it to have a CTEM-like transfer function which is modulated by an envelope that depends on the convergence angle and which has an information limit of twice the probe semi-convergence angle (similar to HAADF or iDPC, and double that of axial BF-STEM). The CTEM-like CTF produces good transfer of low spatial frequencies but the oscillatory behavior complicates analysis at high spatial frequencies and information is lost at the zero crossings.

In this work, we present an analysis of the information transfer in defocused 4D-STEM, including additional coherent and incoherent scattering components that extend beyond the WPOA. The analysis highlights that even under the WPOA, tcBF and the older direct ptychographies of SSB and iDPC draw their information from different parts of a 4D-STEM data set: tcBF sums only the symmetric components of the complex phase contrast transfer function; while SSB and iDPC relies on the antisymmetric components.  

We show that phase information complementary to tcBF can be retrieved from summing over the anti-symmetric components, producing a tilt-corrected DPC (tcDPC) image, which has a cosine form PCTF. Since both symmetric and anti-symmetric mechanisms are transferred in the same dataset, we propose a new phase imaging method that utilizes both channels, which we term aberration-corrected bright-field STEM imaging (acBF-STEM). AcBF correctly combines the two contributions to yield a continuously nonzero CTF that spans from 0 to twice the probe semi-convergence angle. Therefore, acBF circumvents information loss at spatial frequencies that encounter zero crossings in tcBF, while recovering the phase contrast transfer at low frequencies neglected by tcDPC, iDPC or SSB. AcBF achieves the maximal possible phase contrast within the bright field disk under the WPOA and thus can be viewed as an optimal direct ptychography and an analytically tractable approximation to iterative ptychography when imaging a weakly scattering object. 

Strongly scattering objects also modify the intensity of the transmitted wave, often due to inelastic or multiple scattering particularly in thick or heavy-element samples, and the contrast this produces requires analysis beyond the WPOA. Here we identify and derive the contribution of the coherent amplitude component to its own amplitude contrast transfer function (ACTF) and establish that contrast can be enhanced beyond the WPOA limit by ensuring constructive summation of the phase and amplitude contrast at specific spatial frequencies. For the incoherent scattering into the dark field region, we numerically show tilt-corrected dark-field imaging (tcDF) can further enhance contrast transfer and holds promise for one-shot depth scattering for strongly scattering objects.

\section{Contrast in 4D-STEM}\label{sec: wpoa_contrast}
The theory of contrast transfer by electron scattering in conventional transmission electron
microscopy (CTEM) has been extensively studied and, by reciprocity, can be mapped to scanning
transmission electron microscopy (STEM) as well if the momentum vectors of the incident and
outgoing waves are interchanged and the flight direction of the electrons is reversed \cite{cowley_image_1969}, which
follows from time reversal symmetry in the case of elastic scattering. For inelastic scattering,
energy losses become energy gains, but they still preserve the same contrast transfer in imaging.
When the objects do not significantly alter the intensity of the incident electron beam, but only
induce a small phase shift ($<<$ 1 rad) in the scattered wave, i.e., the weak phase object
approximation (WPOA), a linear phase contrast transfer function (PCTF) can be derived.
Alternatively, this can be shown by applying the first-order Born approximation to the scattering
amplitude with the curvature of the Ewald sphere neglected.

\begin{figure}
    \centering
    \includegraphics[width=1\linewidth]{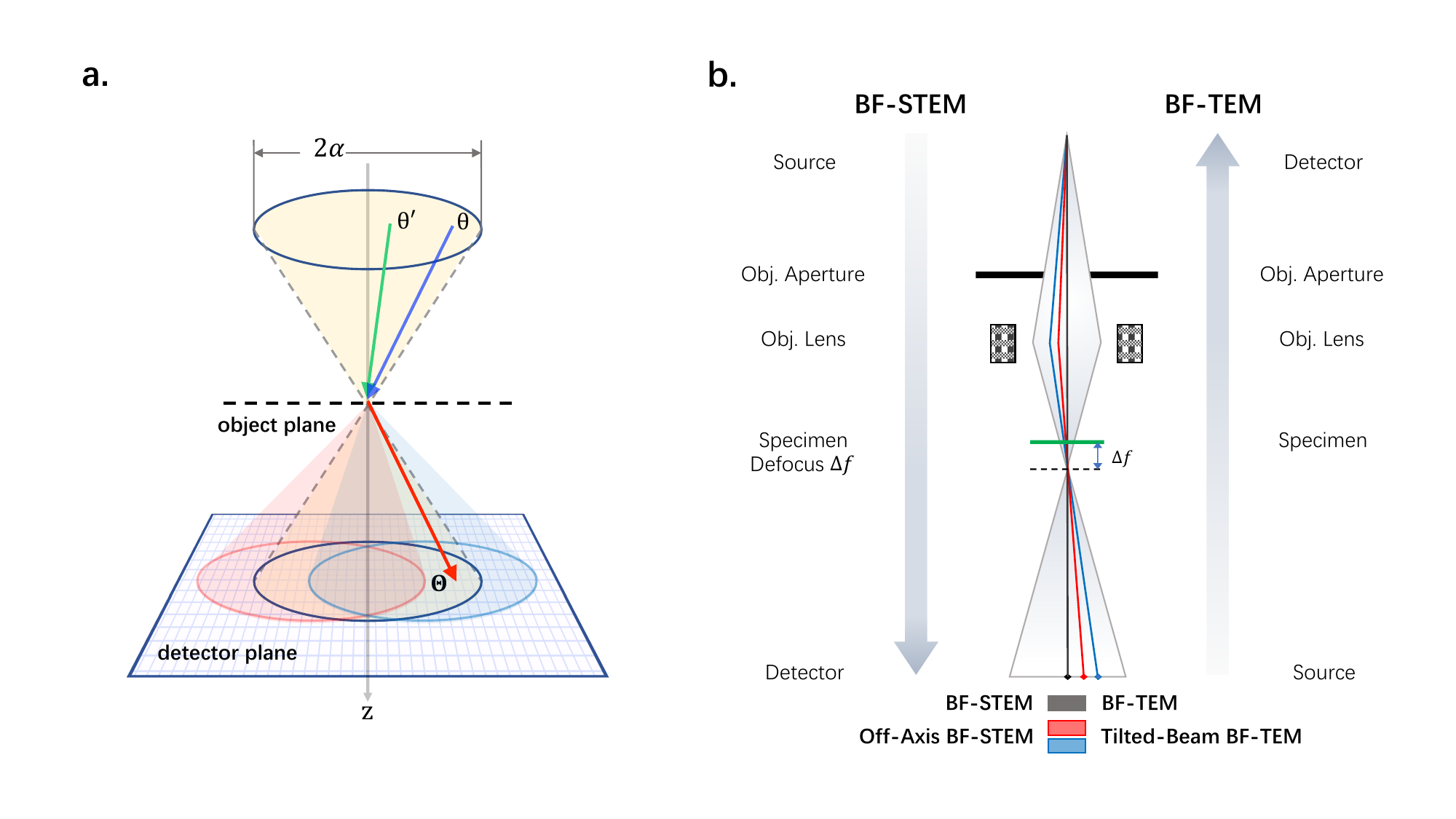}
    \caption{CTEM/STEM reciprocity. (a) Scattering schematic of incident electrons at incident angles $\bm{\theta}$ for the STEM. Scattered electrons at angle $\bm{\theta}$ are recorded by a pixelated detector. (b) Reciprocity of CTEM and STEM. Defocus leads to a shift in images formed with tilted illumination in CTEM or from off-axial BF detectors in STEM.}
    \label{fig:ctem_stem_reciprocity}
\end{figure}

In a four-dimensional scanning transmission electron microscopy 4D-STEM experiment, the two-dimensional convergent beam electron diffraction (CBED)
pattern is recorded by a pixelated detector \cite{tate2016high,philipp2022very} at each probe position in a two-dimensional scan of STEM, as
shown in Figure \ref{fig:ctem_stem_reciprocity}a. The resulting dataset can be interpreted as a collection of two-dimensional 
images with each pixel of the detector acting as a small STEM detector. Conventional imaging
modes, e.g., bright field, annular dark field, differential phase contrast, etc., can be achieved by
summing and subtracting the signals from the appropriate detector pixels accordingly. By
reciprocity \cite{cowley_image_1969}, each image formed from a single detector pixel collected in STEM mode is equivalent
to a plane wave image in CTEM mode with tilted illumination, as shown in Figure \ref{fig:ctem_stem_reciprocity}b.

In quantum theory, scattering is treated as a time-dependent perturbation after an incident plane wave $e^{i \boldsymbol{k} \cdot \mathbf{r}}$ interacts with some scattering potential $V(r)$ \cite{sakurai2020modern}. In the far field limit ($\mathrm{r} \rightarrow \infty$), the asymptotic form of the outgoing wave is
\begin{equation}\label{out_wave}
    \psi^{(+)}(\mathbf{r}) \sim e^{i \boldsymbol{k} \cdot \mathbf{r}}+F\left(\boldsymbol{k}^{\prime}, \boldsymbol{k}\right) \frac{e^{i \boldsymbol{k}^{\prime} \cdot \mathbf{r}}}{r},
\end{equation}
where $\psi^{(+)}(\mathbf{r})$ solves the Lippmann-Schwinger Equation and $F\left(\boldsymbol{k}^{\prime}, \boldsymbol{k}\right)$ is the scattering amplitude which relates to the scattering potential by
\begin{equation}
    F\left(\boldsymbol{k}^{\prime}, \boldsymbol{k}\right)=-\frac{m}{2 \pi \hbar^2}\left\langle\boldsymbol{k}^{\prime}\right| V\left|\psi^{(+)}\right\rangle
    \label{F_kk}.
\end{equation}

Repeated substitution of Equation \ref{F_kk} into \ref{out_wave} leads to a power series expansion in $\psi^{(+)}(\mathbf{r})$. The most commonly used approximation is the first-order Born approximation by taking only the first term in Equation \ref{out_wave}, namely $\left|\psi^{(+)}\right\rangle=|\mathbf{k}\rangle$. In this case, the scattering amplitude is just the three-dimensional Fourier transform of the scattering potential with respect to $\boldsymbol{q}=\boldsymbol{k}-\boldsymbol{k}^{\prime}$.
\begin{equation}\label{F_kk_1_born}
    F^{(1)}\left(\boldsymbol{k}^{\prime}, \boldsymbol{k}\right)=-\frac{m}{2 \pi \hbar^2}\left\langle\boldsymbol{k}^{\prime}\right| V|\boldsymbol{k}\rangle=-\frac{m}{2 \pi \hbar^2} \int e^{-i \boldsymbol{k}^{\prime} \cdot \mathbf{r}} V(\mathbf{r}) e^{i \boldsymbol{k} \cdot \mathbf{r}} d^3 r.
\end{equation}

However, the first order Born approximation is insufficient for calculating all the electron scattering events, as it yields a real-valued scattering amplitude and accounts for only elastic scattering. To conserve the total probability, we must start from the general form in Equation \ref{F_kk} which includes multiple scattering hidden in the higher orders. Rose's formalism \cite{rose_nonstandard_1976} for image contrast provides such a treatment, so we will use it as the basis for our analysis, and we will briefly review it here. Rose gives the time-averaged current density $\overline{J_I}$ to be
\begin{equation}\label{current}
\resizebox{\linewidth}{!}{$
\begin{gathered}
 \overline{J_I}(\boldsymbol{\rho}, \boldsymbol{\Theta})=A(\boldsymbol{\Theta})\left\{1-\frac{2}{\lambda} \mathfrak{Im} \int \mathrm{e}^{-\mathrm{i}(\chi(\bm{\theta})-\chi(\boldsymbol{\Theta}))} \mathrm{e}^{i\left(k(\bm{\theta}-\bm{\Theta}) \cdot \boldsymbol{\rho}+\frac{k z_0\left(|\bm{\theta}|^2-|\boldsymbol{\Theta}|^2\right)}{2}\right)} F(\bm{\theta}, \boldsymbol{\Theta}) \mathrm{d} \Omega_{\bm{\theta}}\right\} \\
 +\frac{1}{k} \sum_{n=0}^{\infty} \frac{k_n}{\lambda_n^2} \iint \mathrm{e}^{-\mathrm{i}\left(\chi_n(\bm{\theta})-\chi_n\left(\bm{\theta}^{\prime}\right)\right)} A(\bm{\theta}) A\left(\bm{\theta}^{\prime}\right) F_n(\bm{\theta}, \boldsymbol{\Theta}) F_n^*\left(\bm{\theta}^{\prime}, \boldsymbol{\Theta}\right) \mathrm{e}^{i\left(k\left(\bm{\theta}-\bm{\theta}^{\prime}\right) \cdot \boldsymbol{\rho}+\frac{k z_0\left(|\bm{\theta}|^2-|\bm{\theta}^{\prime}|^2\right)}{2}\right)} \mathrm{d} \Omega_{\bm{\theta}} \mathrm{d} \Omega_{\bm{\theta}^{\prime}},
\end{gathered}
$}
\end{equation}

where $\bm{\theta}$ is the incident angle and $\boldsymbol{\Theta}$ is the collection angle in STEM (i.e., pixel) on the detector as shown in Figure 1a, $A(\boldsymbol{\Theta})$ is the aperture function equal to 1 inside the aperture $|\bm{\Theta}|<\alpha$, and 0 outside, and $\boldsymbol{\rho}$ is the position in the image. $\chi$ is the phase shift of the scattered electrons attributable to the lens aberrations. The aperture and tilt angles commonly used in electron microscopy are very small, thus the scattering amplitudes $F_n\left(\mathbf{k}_{\mathbf{n}}, \mathbf{K}\right) \approx F_n(\bm{\theta}, \boldsymbol{\Theta})$ describe the angular distribution of the scattered electron waves, where $F=F_{n=0}=F^{(1)}$ is the elastic scattering amplitude, and $n>0$ describes scattering to inelastic channels.

The intensity in the CTEM image with finite angular spread of the illumination may be written as
\begin{equation}\label{equ:intensity}
    I(\boldsymbol{\rho})=\frac{1}{\Omega_0} \int D(\boldsymbol{\Theta}) j(\boldsymbol{\rho}, \boldsymbol{\Theta}) \mathrm{d}^2 \boldsymbol{\Theta},
\end{equation}
where $D(\boldsymbol{\Theta})$ is the illumination function in CTEM, or by reciprocity, the detector response function in STEM, and $\Omega_0$ is the solid angle of the incident beam.

By convention, the image contrast is defined as
\begin{equation}\label{contrast_def}
    C(\boldsymbol{\rho})=1-\frac{I(\boldsymbol{\rho})}{I_0},
\end{equation}
where $I_0$ denotes the average background intensity. In this definition, when there is no specimen scattering the electrons the contrast is zero. Inserting the intensity expression of Equation \ref{equ:intensity} we obtain the contrast\footnote{This expression is equivalent to Equation 26 of Rose \cite{rose_nonstandard_1976}, with the terms grouped differently.} as
\begin{equation}\label{contrast_components}
\resizebox{\linewidth}{!}{$
\begin{gathered}
C(\boldsymbol{\rho})=
\underbrace{\frac{2}{\lambda \Omega_0} \mathfrak{Im} \iint 
A(\bm{\theta}) D(\boldsymbol{\Theta}) A(\bm{\theta}) 
e^{i\left[\chi(\boldsymbol{\Theta})-\chi(\bm{\theta})
+k(\bm{\theta}-\bm{\Theta})\cdot\boldsymbol{\rho}
+\frac{k z_0(|\bm{\theta}|^2-|\Theta|^2)}{2}\right]} 
F_{\mathrm{s}}(\bm{\theta}, \boldsymbol{\Theta}) 
\,d^2\bm{\theta}\,d^2\boldsymbol{\Theta}}_{\text{coherent phase}} \\
+\underbrace{\frac{2}{\lambda \Omega_0} \mathfrak{Re} \iint 
A(\bm{\theta}) D(\boldsymbol{\Theta}) A(\boldsymbol{\Theta}) 
e^{i\left[\chi(\boldsymbol{\Theta})-\chi(\bm{\theta})
+k(\bm{\theta}-\bm{\Theta})\cdot\boldsymbol{\rho}
+\frac{k z_0(|\bm{\theta}|^2-|\Theta|^2)}{2}\right]} 
F_{\mathrm{a}}(\bm{\theta}, \boldsymbol{\Theta}) 
\,d^2\bm{\theta}\,d^2\boldsymbol{\Theta}}_{\text{coherent amplitude}} \\
-\underbrace{\frac{1}{4 \pi^2 \Omega} \mathfrak{Re}
\sum_{n=0}^{\infty} \frac{k_n}{k} k_n^2 \iint 
A(\bm{\theta}) D(\boldsymbol{\Theta}) A(\bm{\theta}^{\prime}) 
e^{i\left[\chi_n(\bm{\theta}^{\prime})-\chi_n(\bm{\theta})
+k_n(\bm{\theta}-\bm{\theta}^{\prime})\cdot\boldsymbol{\rho}
+\frac{k_n z_0(|\bm{\theta}|^2-|\bm{\theta}^{\prime}|^ 2)}{2}\right]} 
F_{{n}}(\bm{\theta}, \bm{\theta}^{\prime}) 
F_{{n}}^*(\boldsymbol{\Theta}, \bm{\theta}) 
\,d^2\bm{\theta}\,d^2\bm{\theta}^{\prime}\,d^2\boldsymbol{\Theta}}_{\text{incoherent amplitude}}.
\end{gathered}
$}
\end{equation}

The complex elastic scattering amplitude is separated into two parts, the Friedel part $F_{\mathrm{s}}$ and the anti-Friedel part $F_{\mathrm{a}}$:
\begin{equation}
    \begin{gathered}
F\left(\boldsymbol{k}^{\prime}, \boldsymbol{k}\right)=F_{\mathrm{s}}\left(\boldsymbol{k}^{\prime}, \boldsymbol{k}\right)+\mathrm{i} F_{\mathrm{a}}\left(\boldsymbol{k}^{\prime}, \boldsymbol{k}\right) \\
F_{\mathrm{s}}\left(\boldsymbol{k}^{\prime}, \boldsymbol{k}\right)=\left[F\left(\boldsymbol{k}^{\prime}, \boldsymbol{k}\right)+F^*\left(\boldsymbol{k}^{\prime}, \boldsymbol{k}\right)\right] / 2 \\
F_{\mathrm{a}}\left(\boldsymbol{k}^{\prime}, \boldsymbol{k}\right)=\left[F\left(\boldsymbol{k}^{\prime}, \boldsymbol{k}\right)-F^*\left(\boldsymbol{k}^{\prime}, \boldsymbol{k}\right)\right] / 2 i.
\end{gathered}
\end{equation}
Both the Friedel part and the anti-Friedel part are contrast-producing. The unitary condition for the scattering amplitude (i.e., that the total current should be preserved) leads to the generalized optical theorem, which connects only the anti-Friedel part of the elastic scattering amplitude with the quadratic terms (i.e. $2^{\text {nd }}$ order in F ) of the elastic and inelastic scattering amplitudes.
\begin{equation}
\resizebox{\linewidth}{!}{$
\begin{gathered}
    F_{\mathrm{a}}=\mathfrak{Im}\left(F\left(\boldsymbol{k}^{\prime}, \boldsymbol{k}\right)\right)=\frac{1}{2 \mathrm{i}}\left\{F\left(\boldsymbol{k}^{\prime}, \boldsymbol{k}\right)-F^*\left(\boldsymbol{k}, \boldsymbol{k}^{\prime}\right)\right\}=\frac{1}{4 \pi} \sum_{n=0}^{\infty} k_{{n}} \int F_{{n}}\left(\boldsymbol{k}^{\prime}, \boldsymbol{k}^{\prime \prime}\right) F_{{n}}^*\left(\boldsymbol{k}^{\prime}, \boldsymbol{k}^{\prime \prime}\right) \mathrm{d} \Omega_{\bm{\theta}^\prime}.
\end{gathered}
$}
\end{equation}

This relation is critical for keeping the consistency of the order of any approximation made, e.g., if the anti-Friedel part is neglected, all other quadratic terms must also be neglected, otherwise the conservation of current is violated.

Inspecting the contrast given by Equation \ref{contrast_components}, the first term describes the phase contrast from the interference between the elastically scattered electrons and the unscattered electrons, and is proportional to the Friedel term of the elastic scattering amplitude $F_{\mathrm{s}}(\bm{\theta}, \boldsymbol{\Theta})=[F(\bm{\theta}, \boldsymbol{\Theta})+ \left.F^*(\boldsymbol{\Theta}, \bm{\theta})\right] / 2$. If the quadratic terms are neglected, the $F_{\mathrm{s}}(\bm{\theta}, \boldsymbol{\Theta})=F(\bm{\theta}, \boldsymbol{\Theta})=F^*(\boldsymbol{\Theta}, \bm{\theta})= F^{(1)}(\bm{\theta}-\boldsymbol{\Theta})$ holds, which is the first order Born approximation that only depends on the scattering vector $\bm{\theta}-\boldsymbol{\Theta}$.

Notice that the second term in Equation \ref{contrast_components}, which is proportional to the anti-Friedel term of the elastic scattering amplitude $F_{\mathrm{a}}(\bm{\theta}, \boldsymbol{\Theta})$, contains the quadratic terms of the scattering amplitudes. It describes the phase contrast from inelastic/plural scattering, and therefore its contribution is proportional to $V^2$ to the lowest possible order of approximation. This term contains interference between electrons scattered from the same illumination angle $\bm{\theta}$ and reaching the same detection angle $\boldsymbol{\Theta}$, regardless of the scattering process. In quantum field theory, these scattering events can be jointly described by the integration in phase space over arbitrary intermediate states $\bm{\theta}^{\prime}$ allowed by Cutkosky's cutting rule \cite{peskin2018introduction}.

The third term in Equation \ref{contrast_components}, also quadratic, is purely incoherent and together with the second term describes scattering absorption contrast, which is the redistribution of amplitude across scattering into the same collection angle $\boldsymbol{\Theta}$ from different illumination angles $\bm{\theta}, \bm{\theta}^{\prime}$. The scattering absorption term has a negative sign, which produces positive contrast, i.e., the object appears as a dark spot on a bright background. The third term is the only one which does not contain $A(\boldsymbol{\Theta})$, which in STEM refers to the objective aperture limit evaluated in the detector space, and thus is the only term able to produce scattering outside the brightfield disk. Therefore, this term entirely captures the dark field signal.
\subsection{Phase Contrast from the Weak Phase Object Approximation (WPOA)}
For weakly scattering objects, the quadratic terms in Equation (7) can be neglected and only the first term remains. Because of the small aperture angles the Ewald sphere can be approximated by a plane, and so we can approximate $\boldsymbol{\theta}^2-\boldsymbol{\Theta}^2 \approx 0$. For a pixelated detector we can assume an infinitely small detector at the location $\boldsymbol{\Theta}_{\boldsymbol{d}}$, and approximate $D(\bm\Theta) \approx \delta\left(\bm{\Theta}-\boldsymbol{\Theta}_{d}\right)$, which eliminates the integral over $\boldsymbol{\Theta}$ (i.e., over the detector extent). See supplementary Figure 1 of reference \cite{yu_dose-efficient_2024} for details on the loss of coherence with finite detector size. In general, we would want to keep each detector angle at more than $3-5 \mathrm{x}$ smaller than the convergence angle. For legibility, we will drop the subscript on the detector pixel location. We define the scattering vector $\boldsymbol{\omega}=\boldsymbol{\theta}-\boldsymbol{\Theta}$ as the difference between the incoming and outgoing wave vectors, which corresponds to the spatial frequency in the sample plane. With these assumptions, Equation \ref{contrast_components} simplifies to,
\begin{equation}\label{contrast_wpoa}
    C_{\mathrm{WPOA}}(\boldsymbol{\rho}, \boldsymbol{\Theta})=\frac{2}{\lambda \Omega_0} A(\boldsymbol{\Theta}) \mathfrak{Im} \iint A(\boldsymbol{\omega}+\boldsymbol{\Theta}) \mathrm{e}^{\mathrm{i}[\chi(\boldsymbol{\Theta})-\chi(\bm{\theta})+k \boldsymbol{\omega} \cdot \boldsymbol{\rho}]} F_{\mathrm{s}}(\boldsymbol{\omega}) \mathrm{d}^2 \boldsymbol{\omega}.
\end{equation}

By taking the inverse Fourier transform of this expression, under WPOA a linear phase contrast transfer function \cite{rose_nonstandard_1976} can be extracted:
\begin{equation}
    \tilde{C}_{\mathrm{WPOA}}(\boldsymbol{\omega}, \boldsymbol{\Theta})=\frac{\operatorname{PCTF}(\boldsymbol{\omega}, \boldsymbol{\Theta}) F_{\mathrm{s}}(\boldsymbol{\omega})}{\lambda},
\end{equation}
\begin{equation}\label{rose_pctf}
\resizebox{\linewidth}{!}{$
\begin{gathered}
\operatorname{PCTF}(\boldsymbol{\omega}, \boldsymbol{\Theta})=\frac{i}{2 \Omega_0} A(\boldsymbol{\Theta})\left\{A(\boldsymbol{\omega}-\boldsymbol{\Theta}) e^{-i[\chi(\boldsymbol{\omega}-\boldsymbol{\Theta})-\chi(-\boldsymbol{\Theta})]}-A(\boldsymbol{\omega}+\boldsymbol{\Theta}) e^{+i[\chi(\boldsymbol{\omega}+\boldsymbol{\Theta})-\chi(\boldsymbol{\Theta})]}\right\},
\end{gathered}
$}
\end{equation}
where we have used the property $F_{\mathrm{s}}(\boldsymbol{\omega})=F_{\mathrm{s}}^*(-\boldsymbol{\omega})$. For the special case of an axial detector ($\boldsymbol{\Theta} =0$), the PCTF reduces to the axial bright field CTF of $-\sin (\chi(\boldsymbol{\omega}))$ with an information cutoff at $|\boldsymbol{\omega}|=\alpha$.

\begin{figure}
    \centering
    \includegraphics[width=\linewidth]{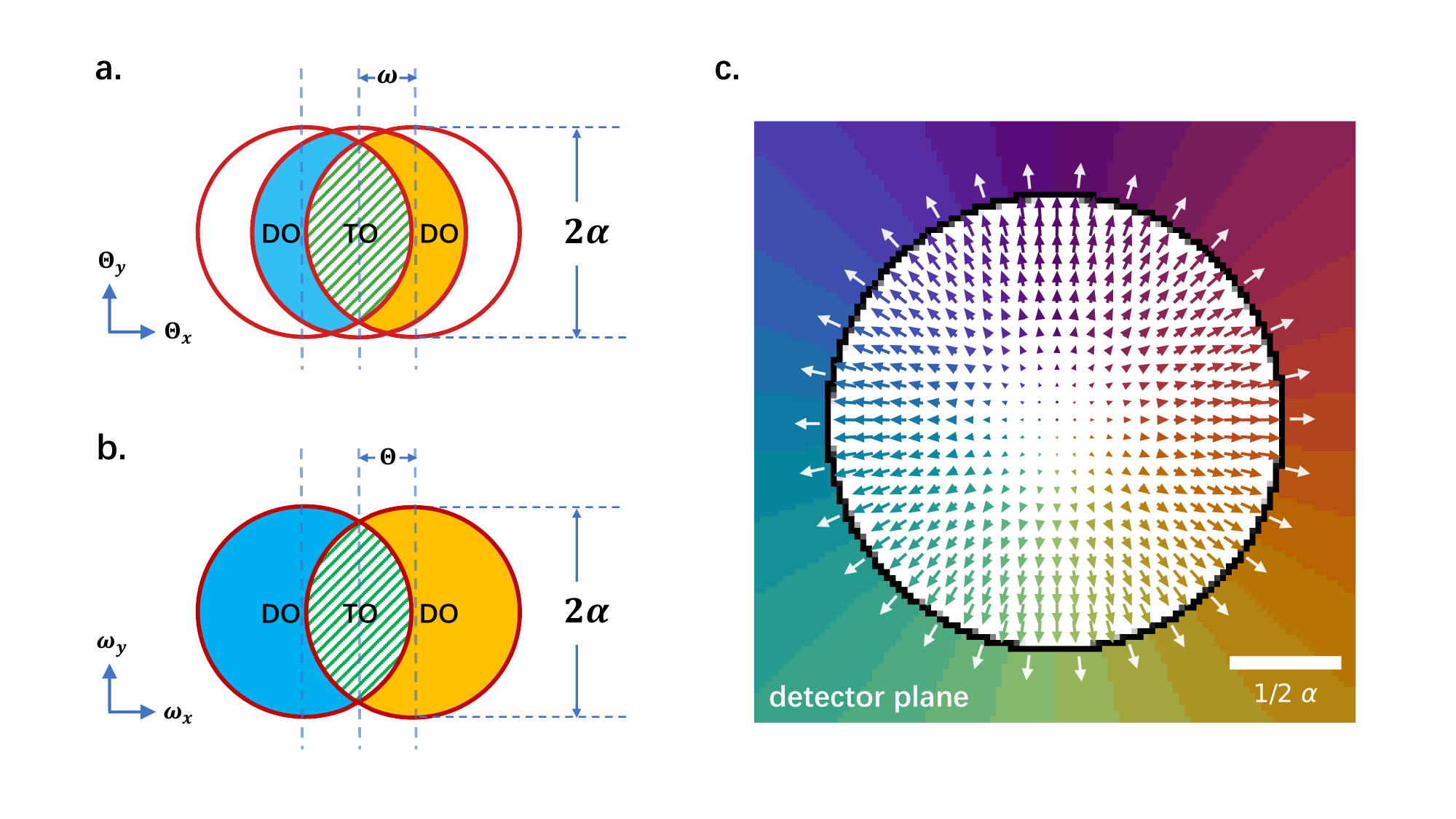}
    \caption{PCTF as a 4D function. (a,b) Projections of the disk overlap regions of the PCTF in the doubly reciprocal space, indicating the double overlap and triple overlap regimes of contrast transfer. (a) When viewed at constant spatial frequency $\boldsymbol{\omega}$, the PCTF shows where information transfer is located in the diffraction plane. (b) When viewed at constant detector position $\boldsymbol{\Theta}$, the PCTF shows the spectral response of an image formed from a single position in the diffraction plane. (c) Schematic showing the real space shifts for the image formed from each detector pixel. Correction of the shifts before combining the images recover the coherent contrast transfer. In tcDF imaging, the shifts are constant within each area of uniform color.}
    \label{fig:info_fig2}
\end{figure}

This PCTF is a four-dimensional function expressed in the doubly-reciprocal space, with $\boldsymbol{\omega}$ representing the spatial frequency response in the Fourier transform of a STEM image from a given location $\bm\Theta$ in diffraction space. Since the virtual STEM images are real (intensity) functions, the CTFs derived here are in all cases conjugate symmetric about $\boldsymbol{\omega}$, though they may take on complex values in many cases. The overlaps of the various aperture terms in Equation (12) delineate two regimes of information transfer, which are shown schematically in Figure 2a and 2b. The region where $A(\boldsymbol{\Theta})=A(\boldsymbol{\omega}-\boldsymbol{\Theta})=A(\boldsymbol{\omega}+\boldsymbol{\Theta})=1$ is referred to as the ``triple overlap'' (TO) and the regions where $A(\boldsymbol{\Theta})=A(\boldsymbol{\omega}-\boldsymbol{\Theta})=1, A(\boldsymbol{\omega}+\boldsymbol{\Theta})=0$ or where $A(\boldsymbol{\Theta})=A(\boldsymbol{\omega}+\boldsymbol{\Theta})= 1, A(\boldsymbol{\omega}-\boldsymbol{\Theta})=0$ are referred to as the ``double overlap'' (DO). When the PCTF is projected along a constant $\boldsymbol{\Theta}$ (Figure \ref{fig:info_fig2}b), the aperture terms form only two circles. The region where they overlap is the TO and where only one is nonzero is the DO. In the direct ptychography literature, the PCTF is typically shown sliced along constant $\boldsymbol{\omega}$, where the figure appears as three circles (Figure \ref{fig:info_fig2}a); the naming of the regions is more natural there, with the TO region corresponding to the overlap of all three circles and the DO region as the areas where only two overlap \cite{pennycook_efficient_2015}.

\subsubsection*{Noise Effects on Information Transfer}
The CTF describes how different spatial frequencies in an object are transferred from the object to the image by the imaging system, assuming an infinite signal-to-noise ratio (SNR). In practice, noise considerations are critical in determining the actual information transfer and thus dose efficiency, particularly when imaging beam sensitive samples. One approach to incorporating noise is by defining the detective quantum efficiency (DQE) \cite{mcmullan_detective_2009},
\begin{equation}\label{dqe_def}
    \operatorname{DQE}(\boldsymbol{\omega})=\operatorname{DQE}(0) \frac{|\operatorname{PCTF}(\boldsymbol{\omega})|^2}{|\operatorname{NPS}(\boldsymbol{\omega})|^2},
\end{equation}
where $\operatorname{DQE}(0)=1$ in a perfect pixel detector and $\operatorname{NPS}(\boldsymbol{\omega})$ is the noise power spectrum (NPS). For an ideal detector the measurement noise is only from Poisson noise. The noise spectrum of a signal with Poisson noise is a nonlinear function that depends on the properties of the underlying signal, but if the signal is nearly uniform (as will be the case for bright-field imaging of a weakly scattering object) then the NPS can be approximated as flat \cite{lucke2001fourier}.

DQE is a valuable metric in that it provides a sample and dose-level independent comparison of different imaging methods.  Bennemann et al \cite{bennemann_detective_2025} derive the DQE for in-focus and defocused SSB. \ref{appendix_part3_C}  extends this to the tilt-corrected methods and shows how different phase retrieval strategies in 4DSTEM can alter the NPS and thus the DQE, even if the CTF is the same. In addition, that the DQE is defined using the modulus squared $|\operatorname{PCTF}(\boldsymbol{\omega})|^2$ allows for its natural extension to nonlinear imaging methods, e.g., iterative ptychography, where the exact CTFs are mathematically intractable and only the power spectrum can be obtained from simulations. However, the nonlinear dependence of the realized noise power spectrum on the underlying signal can pose challenges when the DQE is evaluated numerically for test objects and makes such results difficult to generalize as they are dose-dependent.

\subsubsection{Tilt-corrected Bright Field (tcBF)}
In experiments, we always desire an aberration-free probe, apart from the defocus which has been shown to provide extra contrast transfer. If we consider only a defocus $\Delta f$, the aberration function is $\chi(\boldsymbol{\Theta})=-\frac{1}{2} k_0 \Delta f |\boldsymbol{\Theta}|^2$, and the phase contrast transfer function becomes
\begin{equation}
\resizebox{\linewidth}{!}{$
\begin{gathered}
    \operatorname{PCTF}(\boldsymbol{\omega}, \boldsymbol{\Theta})=\frac{i}{2 \Omega_0} A(\boldsymbol{\Theta})\left\{A(\boldsymbol{\omega}-\boldsymbol{\Theta}) e^{+1 / 2 i k_0 \Delta f |\boldsymbol{\omega}|^2}-A(\boldsymbol{\omega}+\boldsymbol{\Theta}) e^{-1 / 2 i k_0 \Delta f |\boldsymbol{\omega}|^2}\right\} e^{-i(\Delta f \boldsymbol{\Theta}) \cdot\left(k_0 \boldsymbol{\omega}\right)}.
\end{gathered}
$}
\end{equation}

It can be shown that $\operatorname{PCTF}(\boldsymbol{\omega}, \boldsymbol{\Theta})$ is conjugate symmetric about $\boldsymbol{\Theta}$:
\begin{equation}\label{pctf_conjugate}
    \operatorname{PCTF}(\boldsymbol{\omega},-\boldsymbol{\Theta})^*=\operatorname{PCTF}(\boldsymbol{\omega},+\boldsymbol{\Theta}).
\end{equation}

The factor $e^{-i(\Delta f \boldsymbol{\Theta}) \cdot\left(k_0 \boldsymbol{\omega}\right)}$ in this PCTF represents, by the Fourier shift theorem, that the image formed from the detector pixel at location $\boldsymbol{\Theta}$ in the diffraction plane is shifted in real space by a distance $\Delta \boldsymbol{\rho}=\Delta f \boldsymbol{\Theta}$ as shown schematically in Figure \ref{fig:ctem_stem_reciprocity}b. In tilt-corrected bright-field STEM, these shifts are removed and the shift-corrected images from all detector pixels within the bright-field disk are summed. These shifts may be sub-pixel with respect to the probe raster scan, enabling recovery of images with finer spatial sampling by upsampling prior to shifting, as demonstrated in Yu et al \cite{yu_dose-efficient_2022}. We use the superscript $^{(\mathrm{tc})}$ to denote the tilt-corrected CTF at each detector position that remains when the shift is removed,
\begin{equation}\label{df_shift}
    \operatorname{PCTF}(\boldsymbol{\omega}, \boldsymbol{\Theta})=\operatorname{PCTF}^{(\mathrm{tc})}(\boldsymbol{\omega}, \boldsymbol{\Theta}) e^{-i(\Delta f \bm{\Theta}) \cdot\left(k_0 \boldsymbol{\omega}\right)}.
\end{equation}
% \textcolor{red}{typo in Theta, fix in proofread}\\

The PCTF of the sum of all shift-corrected images within the bright-field disk is
\begin{equation}
    \begin{aligned}
\operatorname{PCTF}_{\mathrm{tcBF}}(\boldsymbol{\omega}) & =\frac{1}{2} \sum_{\boldsymbol{\Theta}} \operatorname{PCTF}^{(\mathrm{tc})}(\boldsymbol{\omega},+\boldsymbol{\Theta})+\operatorname{PCTF}^{(\mathrm{tc})}(\boldsymbol{\omega},-\boldsymbol{\Theta})\\
&=\sum_{\boldsymbol{\Theta}} \Re \mathrm{e}\left(\operatorname{PCTF}^{(\mathrm{tc})}(\boldsymbol{\omega}, \boldsymbol{\Theta})\right) \\
& =-\mathbb{A}(\boldsymbol{\omega}) \sin \left(1/2 k_0 \Delta f |\boldsymbol{\omega}|^2\right).
\end{aligned}
\end{equation}

Here $\mathbb{A}(\boldsymbol{\omega})$ is the normalized area of the overlap region between two disks separated by spatial frequency $\omega \in[0,2 \alpha]$, which we derive in \ref{appendix_part3_C}. TcBF thus can have an information limit up to $2 \alpha$, doubling that of axial BF-STEM. An example of this contrast transfer function is shown in Figure \ref{fig: info_fig3_ctf_dqe}a.

Note that the appearance of the shift term in Equation \ref{rose_pctf} is not unique to the defocus aberration. The image shift of an off-axial STEM image is given by Lupini et al \cite{lupini_rapid_2016} as $\Delta \boldsymbol{\rho}(\boldsymbol{\Theta})=\nabla \chi({\boldsymbol{\Theta}})$. This can be derived by considering a Taylor expansion of Equation (12) with respect to $\chi$, or via the Wigner-Weyl transform as shown by Ma et al \cite{cueva2021transforming, ma_emittance_2025}. However, the property \ref{pctf_conjugate} does not hold for all aberrations, so it is not generally the case that the tcBF PCTF will be the corresponding CTEM PCTF multiplied by the envelope function $\mathbb{A}(\boldsymbol{\omega})$.

\subsubsection{Tilt-corrected Differential Phase Contrast (tcDPC) }
We now consider the general case when defocus is present and explore the benefit of other combinations of the shift-corrected images. In fact, shift-correction of DPC images in samples thicker than the depth of field has already been shown to help refocus the DPC image and obtain defocus estimates \cite{seifer_flexible_2021}. Since the PCTF contains both a symmetric and antisymmetric contribution, with the symmetric term leading to tcBF, we now explore the effect on the antisymmetric term. 
%At zero defocus, the antisymmetric term would lead to DPC or SSB imaging and at finite defocus the image would be additionally blurred by the defocus shifts.

The conjugate symmetry of the PCTFs with respect to opposing detector positions (Equation \ref{pctf_conjugate}) suggests that summing the tilt-corrected images leads to cancellation of terms in the double overlap region (where the PCTF is complex-valued) that suppresses contrast that is present in the individual images. Subtracting the shift-corrected images from opposite detector positions $\boldsymbol{\Theta}$, $-\boldsymbol{\Theta}$ extracts the phase contrast transfer in the double overlap region and is equivalent to summing $\operatorname{PCTF}(\boldsymbol{\omega}, \boldsymbol{\Theta})$ over all $\boldsymbol{\Theta}$ anti-symmetrically. We term this contribution as tilt-corrected DPC (tcDPC). Its transfer function is
\begin{equation}
\begin{aligned}
\operatorname{PCTF}_{\mathrm{tcDPC}}(\boldsymbol{\omega}) & =\frac{1}{2} \sum_{\boldsymbol{\Theta}} \operatorname{PCTF}^{(\mathrm{tc})}(\boldsymbol{\omega},+\boldsymbol{\Theta})-\operatorname{PCTF}^{(\mathrm{tc})}(\boldsymbol{\omega},-\boldsymbol{\Theta})\\
&=i \sum_{\boldsymbol{\Theta}} \Im \mathfrak{m}\left(\operatorname{PCTF}^{(\mathrm{tc})}(\boldsymbol{\omega}, \boldsymbol{\Theta})\right) \\
& =i\big (\mathbb{A}(\boldsymbol{\omega})-\mathbb{A}(2 \boldsymbol{\omega})\big) \cos \left(1 / 2 k_0 \Delta f |\boldsymbol{\omega}|^2\right).
\end{aligned}
\end{equation}
The tcDPC envelope is zero at zero spatial frequency and at $2 \alpha$, reaching a maximum of 0.41 at $0.9 \alpha$. Cancellation of terms in the triple overlap region (where the PCTF is real) leads to the loss of low spatial frequency information. The cosine modulation implies that the optimal CTF is achieved at in-focus conditions with no aberrations inside the aperture, which reduces tcDPC to conventional in-focus DPC. An example PCTF for tcDPC imaging is shown in Figure \ref{fig: info_fig3_ctf_dqe}a.

The PCTF of tcDPC is purely imaginary, producing a gradient or differential phase image in real space, as in conventional DPC. Integration of the differential signal, termed i-tcDPC, yields the phase image\footnote{Practical computation of the i-tcDPC image would proceed by computing the $\mathrm{tcDPC}_{\mathrm{x}}$ and $\mathrm{tcDPC}_{\mathrm{y}}$ components by subtracting opposing detector images across each axis to produce the two vector components of the gradient which are needed for the 2D integration, but this does not affect the information transfer.}. Integration in real space corresponds to a division by the imaginary unit times spatial frequency in Fourier space,
\begin{equation}
    \operatorname{PCTF}_{\mathrm{i}-\mathrm{tcDPC}}(\boldsymbol{\omega})=\frac{\operatorname{PCTF}_{\mathrm{tcDPC}}(\boldsymbol{\omega})}{i \boldsymbol{\omega}}.
\end{equation}
The above transformation performed in the integration procedure will affect the shape of the PCTF (particularly by boosting low spatial frequencies) but we emphasize that this does not increase the information content of the reconstructed image as the measurement noise is similarly transformed. The division by $i \boldsymbol{\omega}$ to transform from tcDPC to i-tcDPC is applied to both the PCTF and the NPS in Equation \ref{dqe_def} and thus cancel out, leaving the DQE equal to the $\operatorname{PCTF}^2$. Typically, a high-pass filter is applied to iDPC images to suppress the low-frequency noise which is amplified by the Fourier integration step \cite{lazic_phase_2016}. Since a pixelated detector is used, the technique may also be referred to as center-of-mass (CoM)/integrated center-of-mass (iCoM) imaging.

\subsubsection{Aberration-corrected Bright Field (acBF)/Direct Ptychography}
The $\sin (\chi(\boldsymbol{\omega}))$ modulation of tcBF and the $\cos (\chi(\boldsymbol{\omega}))$ modulation of tcDPC (along with their different envelope functions) suggest that an optimally efficient imaging method can be produced by appropriately combining the images formed from each method, especially since zero crossings in tcBF will be where tcDPC is maximal and vice versa. A direct approach, whereby the two images are first independently formed by summation and subtraction of the virtual STEM images and then combined as $\operatorname{I}_{\mathrm{tcBF}}(\boldsymbol{\omega})-i\operatorname{I}_{\mathrm{tcDPC}}(\boldsymbol{\omega})$, is discussed in \ref{appendix_part3_A}. This approach is intuitively appealing but, in our experience, suffers from numerical challenges in implementation.

An alternative approach is to apply a complex Fourier filter to each shift-corrected STEM image which corrects for the phase of the PCTF at each spatial frequency, prior to summation. For each shift-corrected image at tilted incidence, we apply a frequency-dependent phase shift to rotate the complex $\operatorname{PCTF}(\boldsymbol{\omega}, \boldsymbol{\Theta})$ at each $\boldsymbol{\omega}$ to become real and positive so that the sum of $\operatorname{PCTF}(\boldsymbol{\omega}, \boldsymbol{\Theta})$ over (2) is maximized as $|x+y| \leq|x|+|y|$. In this way, the total phase contrast transfer $\operatorname{PCTF}(\boldsymbol{\omega})$ equals the sum of the absolute value of the $\operatorname{PCTF}(\boldsymbol{\omega}, \boldsymbol{\Theta})$ at each $(\boldsymbol{\omega}, \boldsymbol{\Theta})$.

The correction factor required depends on the region in frequency space. Specifically, for each of the disk overlap regions, the correction factors and the resulting corrected PCTF values are:

\textbullet\ If $|\boldsymbol{\omega}+\boldsymbol{\Theta}|>\alpha$ and $|\boldsymbol{\omega}-\boldsymbol{\Theta}|<\alpha$, apply complex phase shift $e^{-1 / 2 i k_0 \Delta f|\boldsymbol{\omega}|^2}$,
\begin{equation}\label{equ:acbf_do1}
    A(\boldsymbol{\Theta})\left\{A(\boldsymbol{\omega}-\boldsymbol{\Theta}) e^{+1 / 2 i k_0 \Delta f |\boldsymbol{\omega}|^2}-A(\boldsymbol{\omega}+\boldsymbol{\Theta}) e^{-1 / 2 i k_0 \Delta f |\boldsymbol{\omega}|^2}\right\} e^{-1 / 2 i k_0 \Delta f |\boldsymbol{\omega}|^2}=1.
\end{equation}

\textbullet\ If $|\boldsymbol{\omega}+\boldsymbol{\Theta}|<\alpha$ and $|\boldsymbol{\omega}-\boldsymbol{\Theta}|>\alpha$, apply complex phase shift $-e^{+1 / 2 i k_0 \Delta f |\boldsymbol{\omega}|^2}$,
\begin{equation}\label{equ:acbf_do11}
\resizebox{\linewidth}{!}{$
\begin{gathered}
A(\boldsymbol{\Theta})\left\{A(\boldsymbol{\omega}-\boldsymbol{\Theta}) e^{+1 / 2 i k_0 \Delta f |\boldsymbol{\omega}|^2}-A(\boldsymbol{\omega}+\boldsymbol{\Theta}) e^{-1 / 2 i k_0 \Delta f |\boldsymbol{\omega}|^2}\right\}\left(-e^{+1 / 2 i k_0 \Delta f |\boldsymbol{\omega}|^2}\right)=1.
\end{gathered}
$}
\end{equation}

\textbullet\ If $|\boldsymbol{\omega}+\boldsymbol{\Theta}|<\alpha$ and $|\boldsymbol{\omega}-\boldsymbol{\Theta}|<\alpha$, only sign flipping is needed as the PCTF is already real in this region,
\begin{equation}\label{equ:acbf_to}
\resizebox{\linewidth}{!}{$
\begin{gathered}
\left|A(\boldsymbol{\Theta})\left\{A(\boldsymbol{\omega}-\boldsymbol{\Theta}) e^{+1 / 2 i k_0 \Delta f |\boldsymbol{\omega}|^2}-A(\boldsymbol{\omega}+\boldsymbol{\Theta}) e^{-1 / 2 i k_0 \Delta f |\boldsymbol{\omega}|^2}\right\}\right|=2\left|\sin \left(k_0 \Delta f |\boldsymbol{\omega}|^2 / 2\right)\right|.
\end{gathered}
$}
\end{equation}

In practice, instead of taking the absolute value, it is straightforward to apply the phase shift: $e^{-1 / 2 i k_0 \Delta f |\boldsymbol{\omega}|^2} e^{-i \tan ^{-1}\left(\frac{\sin \left(k_0 \Delta f |\boldsymbol{\omega}|^2\right)}{1+\cos \left(k_0 \Delta f |\boldsymbol{\omega}|^2\right)}\right)}$ as discussed in \ref{appendix_part3_B},
\begin{equation}
\label{equ:acbf_all}
\resizebox{\linewidth}{!}{$
\begin{gathered}
A(\boldsymbol{\Theta})\left\{A(\boldsymbol{\omega}-\boldsymbol{\Theta}) e^{+1 / 2 i k_0 \Delta f |\boldsymbol{\omega}|^2}-A(\boldsymbol{\omega}+\boldsymbol{\Theta}) e^{-1 / 2 i k_0 \Delta f |\boldsymbol{\omega}|^2}\right\} e^{-1 / 2 i k_0 \Delta f |\boldsymbol{\omega}|^2} e^{-i \tan ^{-1}\left(\frac{\sin \left(k_0 \Delta f |\boldsymbol{\omega}|^2\right)}{1+\cos \left(k_0 \Delta f |\boldsymbol{\omega}|^2\right)}\right)} \\
=\left|1-e^{-i k_0 \Delta f |\boldsymbol{\omega}|^2}\right|=2 \sqrt{\sin ^2\left(k_0 \Delta f |\boldsymbol{\omega}|^2 / 2\right)}.
\end{gathered}
$}
\end{equation}

Summing the corrected transfer functions over all $\boldsymbol{\Theta}$ gives the full PCTF,
\begin{equation}\label{equ: acbf_ctf}
    \operatorname{PCTF}_{\mathrm{acBF}}(\boldsymbol{\omega})=[\mathbb{A}(\boldsymbol{\omega})-\mathbb{A}(2 \boldsymbol{\omega})]+\mathbb{A}(2 \boldsymbol{\omega})\left|\sin \left(k_0 \Delta f |\boldsymbol{\omega}|^2 / 2\right)\right|.
\end{equation}

\begin{figure*}[h]
\centering
\includegraphics[width=\linewidth]{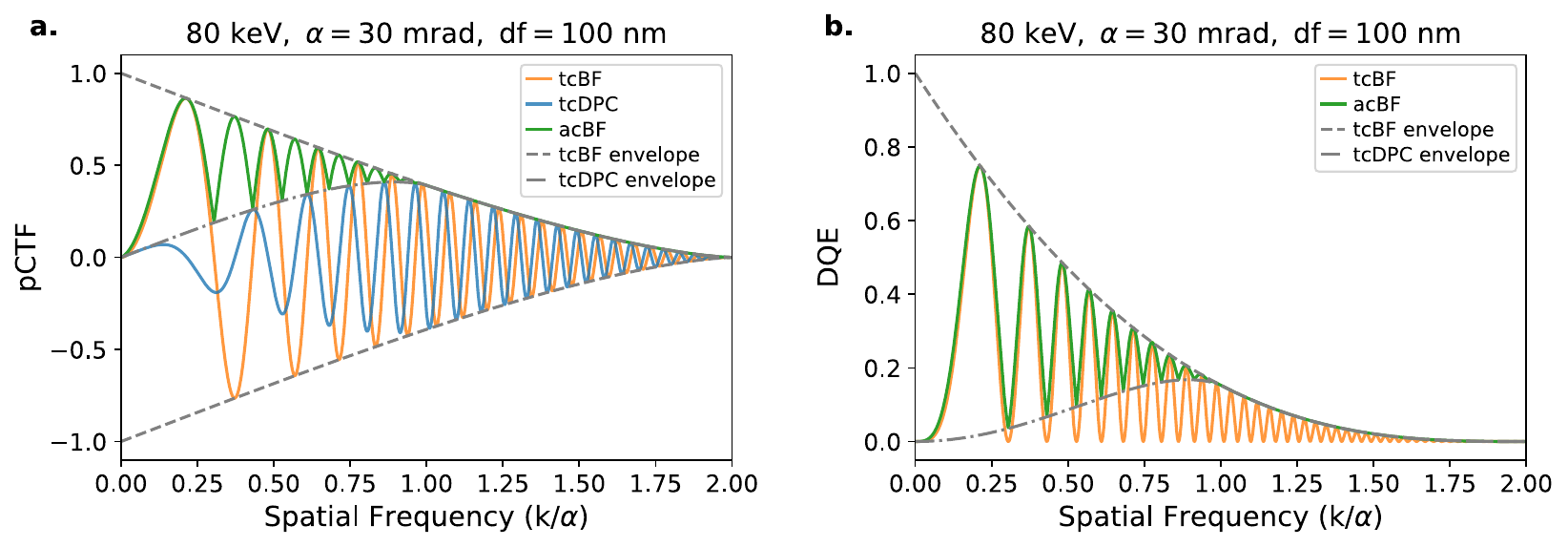}
\caption{ (a) Phase contrast transfer function (PCTF) and (b) Detective quantum efficiency (DQE)
of tcBF, tcDPC and acBF at 300 kV, 30 mrad convergence semi-angle. The tcDPC curve in (a) shows the imaginary part of its PCTF. The
tcDPC envelope is equal to the PCTF of in-focus tcDPC. acBF fills in the difference between tcBF and tcDPC, which yields constantly non-zero contrast transfer.} 
\label{fig: info_fig3_ctf_dqe}
\end{figure*}

AcBF yields a PCTF and DQE that combines the best of both tcBF and tcDPC. As shown in Figure \ref{fig: info_fig3_ctf_dqe}, acBF (i) gains significant information surplus from 0 to $\alpha$ by utilizing the triple overlap region neglected by tcDPC or SSB; and (ii) produces continuous nonzero contrast transfer from 0 to $2 \alpha$, eliminating information loss at any frequency due to the zero crossings in tcBF. Figure 3b visualizes the significant information surplus the triple overlap region can contribute by comparing the DQE of in-focus iDPC and SSB with defocused tcBF and acBF . AcBF with defocused data provides detective efficiency that is at least as high as the other methods at every spatial frequency, and is substantially higher overall at frequencies between 0 and $\alpha$.

\begin{figure*}[h]
\centering
\includegraphics[width=\linewidth]{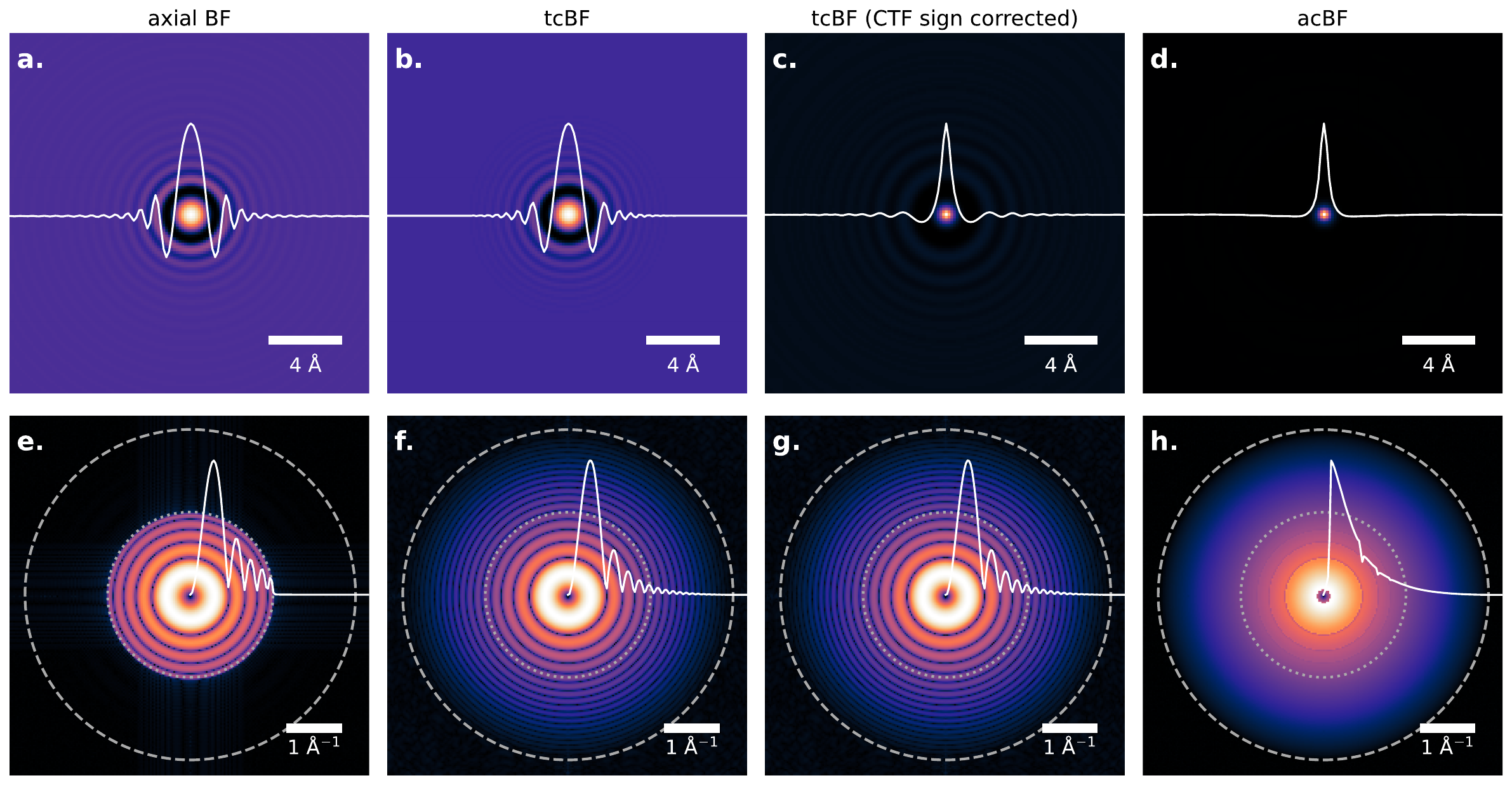}
\caption{Imaging of a simulated single C atom at 300 keV and 30 mrad convergence semi-angle under different methods, (a) axial BF, (b) tcBF, (c) tcBF with post-summation CTF correction (sign flipping), (d) acBF. (e-h) Corresponding fast Fourier transform (FFT) power spectra of (ad). The dashed circles on the FFTs indicate the spatial frequencies corresponding to $1 \alpha$ and $2 \alpha$. acBF demonstrates superior information transfer with no information loss at all frequencies up to $2 \alpha$.} 
\label{fig: info_fig4_Catom}
\end{figure*}

In Figure \ref{fig: info_fig4_Catom} we compare the performance of acBF to other techniques on a multislice simulation of a single atom and obtain an approximate point spread function. Compared to the ordinary axial BF image, the tcBF point spread function has similar oscillatory tails but, due to the $2 \alpha$ information limit, has a sharper central lobe. Phase contrast image interpretability in axial $\mathrm{BF} / \mathrm{tcBF}$ can be improved by ``CTF sign-correction'', which simply inverts the contrast at spatial frequencies where the PCTF is negative in the \textit{summed image}, as commonly used in cryo-EM \cite{downing2008restoration}. This correction suppresses the intensity in the tails but does not eliminate them, as the zero-crossings in the PCTF are still present. AcBF provides a sharp central lobe with no visible tails and no zeroes in the PCTF, which will produce interpretable phase images from complex structures at high resolution.

We applied the acBF reconstruction procedure to experimental data \cite{li2025atomically} of a radiation-sensitive and weakly-scattering NU-1000 metal-organic framework (MOF) acquired at low dose as shown in Figure 5, and we observe that acBF gives the same information limit and ability to resolve single atoms in the linker nodes as multislice electron ptychography. Within the WPOA, acBF is maximally efficient in the limited sense of Dwyer and Paganin \cite{dwyer_quantum_2024} for 4D-STEM without a phase plate or diffuser, and so represents the upper limit of information transfer for STEM phase retrieval methods within the bright field disk ($2\alpha$) within that approximation. For thin samples in the low dose regime where practically no electrons scatter to the dark field region, $\operatorname{acBF}$ also sets the upper limit for iterative ptychography when imaging a weak phase object. We will see below there is additional usable information that is still neglected by the WPOA that is accessible to iterative ptychography methods.

\begin{figure*}[h]
\centering
\includegraphics[width=\linewidth]{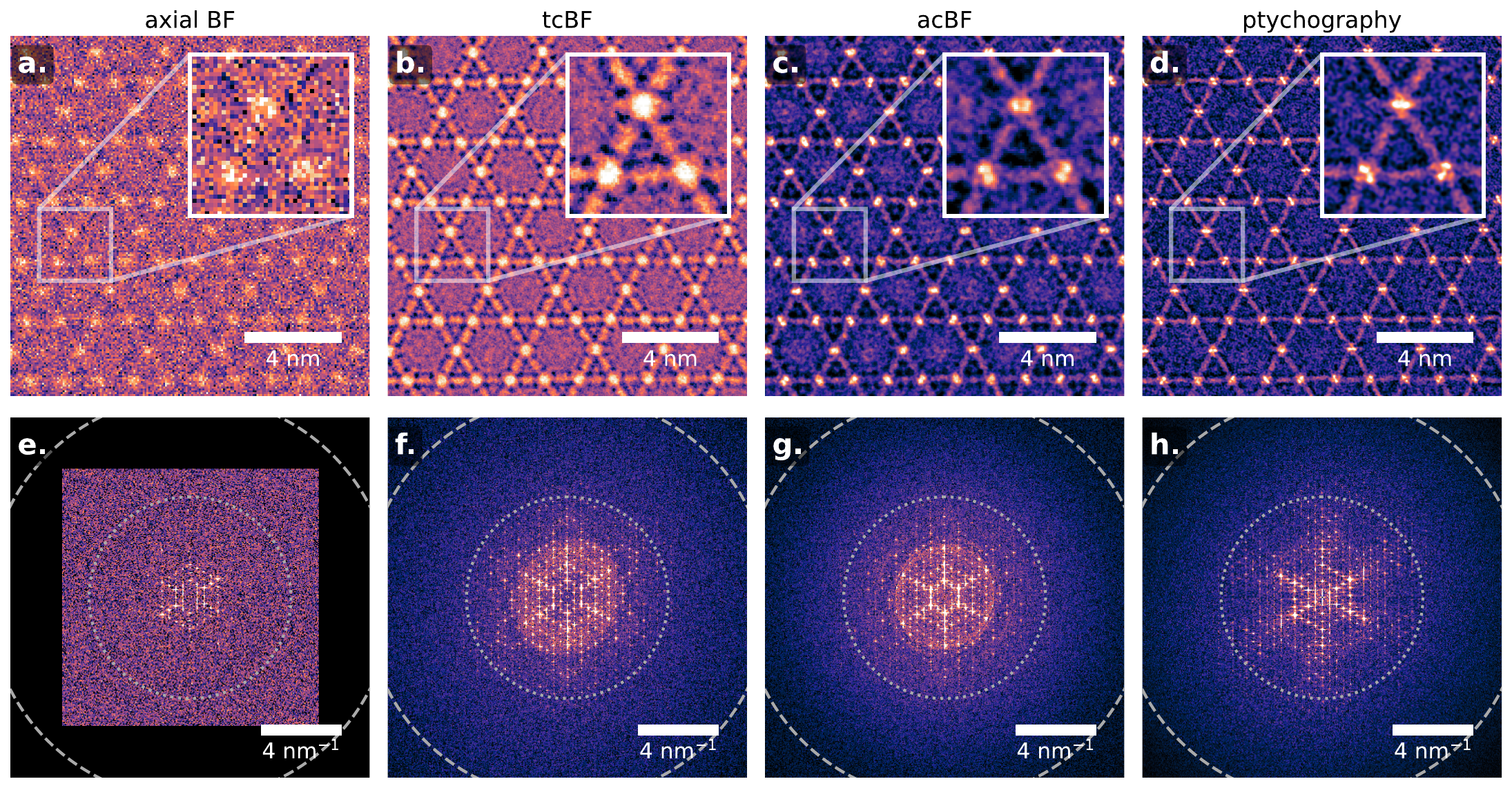}
\caption{Imaging of a dataset (taken from \cite{li2025atomically}) of a 48-nm-thick metal-organic framework acquired at a dose of $\sim 100 \mathrm{e-} / \text{Å}^2$ reconstructed using different 4D-STEM phase contrast methods, (a) axial BF, (b) tcBF, (c) acBF, (d) multislice ptychography, along with their respective Fourier transforms (e-h). tcBF and acBF have been upsampled by 2. A zoom-in of the NU-1000 structure shows the Zr clusters are well resolved by both acBF and ptychography, and the Fourier transforms $(\mathrm{g}-\mathrm{h})$ indicates a resolution of better than $2 \text{Å}$. The dashed circles on the FFTs indicate the spatial frequencies corresponding to $1 \alpha$ and $2 \alpha$.} 
\label{fig: info_fig5_MOF}
\end{figure*}

There is a connection between acBF and aberration-corrected SSB ptychography. The SSB approach begins from the same 4-dimensional PCTF object as we consider here, but its algorithm considers each image spatial frequency and chooses what information in the diffraction space to sum, rather than our approach which considers each diffraction position and transforming image spatial frequencies. For in-focus data, the SSB approach offers a more natural approach for rejecting the measurement noise in regions of the data that have no information transfer and so can result in a higher DQE. (This is discussed in more detail in \ref{appendix_part3_C} and \cite{bennemann_detective_2025}). The advantage of the tilt-corrected imaging approach is more apparent when dealing with defocused data. First, the noise rejection advantage of SSB disappears because there is useful information transfer in both the DO and TO regions. In addition, in the tilt-correction methods the defocus manifests as image shifts, which are computationally simple to detect and correct using cross correlation and the Fourier shift theorem. This also produces a reliable measurement of the defocus, which is used to generate the correction factors applied to each virtual STEM image. In the SSB approach, by contrast, handling defocus requires fitting a plane through a complex-valued function within the DO "trotters" and suffers from challenges due to the $2 \pi$ wrapping of the values. This difference makes the tilt-corrected family of methods more straightforward to apply to defocused data, with the resulting advantage that it becomes feasible to utilize the strong information transfer at low spatial frequencies that we have shown are present when defocus is used. As mentioned previously, the tilt-correction procedure also allows for sub-pixel shifts and therefore a reconstruction in real space at a finer spatial sampling than the real-space distances between recorded diffraction patterns as in iterative ptychography (compare Figure \ref{fig: info_fig5_MOF}e v.s. f-h).

\subsection{Contrast from the Weak Amplitude Object Approximation (WAOA)}
So far, analysis of the contrast transfer has only treated the weak phase object response---both tcBF and tcDPC are described by the first term of Equation \ref{contrast_components}. Here we consider the contributions from the next term. Scattering by real samples also leads to a damping of the intensity of the collected electron wave (often termed ``absorption,'' though the high energy electrons in a TEM are rarely truly absorbed by the sample). Change in the intensity of the transmitted beam is caused mainly by scattering outside the range of collected angles. In wave optics, this is often modeled by a complex scattering potential under the WKB/eikonal approximation \cite{reimer2013transmission}, which will produce an attenuation term in the transmission function. However, it is worth noting that the WKB approximation does not always obey the optical theorem in the quantum theory of scattering. A more rigorous treatment originating from the optical theorem shows that part of the intensity loss from the incident beam is intrinsically linked to the imaginary part of the elastic scattering amplitude, which is coherent and preserves phase information, while the remainder is incoherent and key for forming the dark field signal.

\begin{figure*}[h]
\centering
\includegraphics[width=\linewidth]{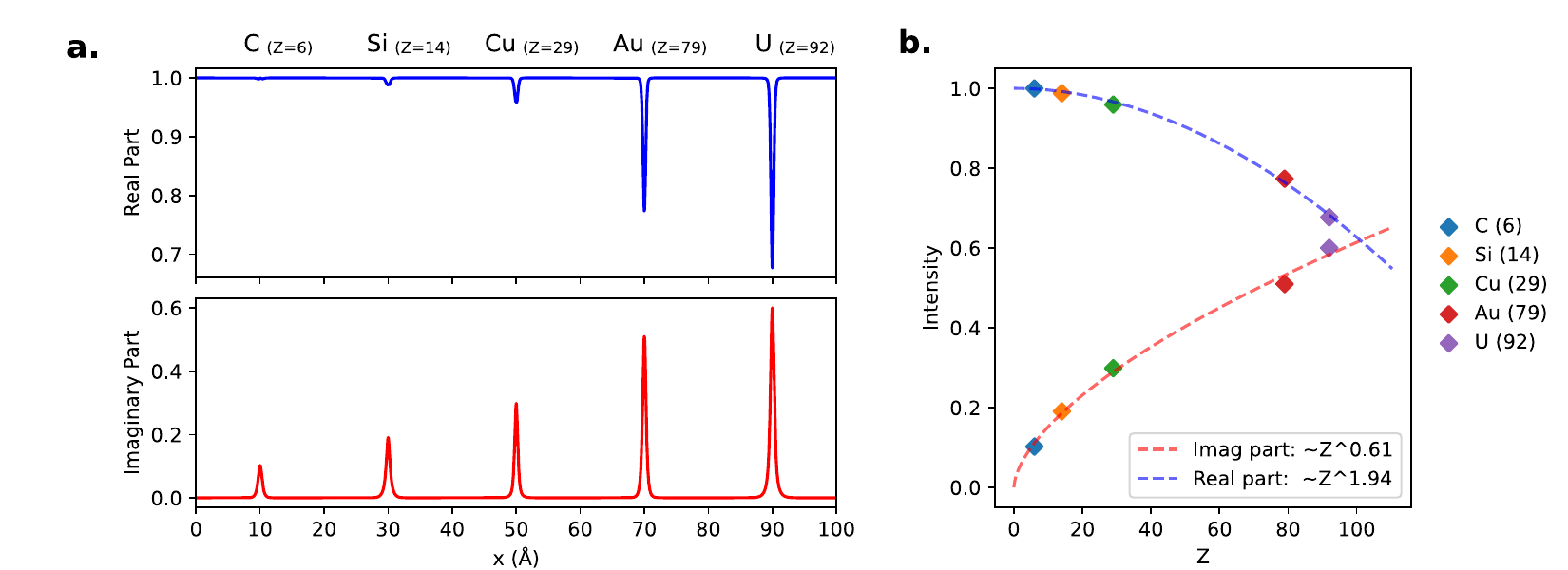}
\caption{(a) Line profile of the complex exit wave for five isolated single atoms $\mathrm{C}, \mathrm{Si}, \mathrm{Cu}, \mathrm{Au}$, U evenly spaced every 20 \AA\ with an incident electron beam energy of 200 keV . The exit waves are calculated using multislice simulations with a real space sampling size of 0.05 \AA\ and slice thickness of 2 \AA\. (b) Different Z-dependence of the real and imaginary parts of the complex exit waves.} 
\label{fig: info_fig6_exitwave}
\end{figure*}

The second term in Equation \ref{contrast_components} contains the anti-Friedel component of the elastic scattering amplitude, which leads to an amplitude change to the current in Equation \ref{current}. Since this component contains quadratic terms of the scattering amplitudes $F_n$, the amplitude contrast transfer will have a different atomic-number (Z)-dependence from phase contrast, as shown for single atoms of several different elements in Figure \ref{fig: info_fig6_exitwave}. The phase change of the scattered wave scales with $Z^{\sim 0.61}$, similar to the observations in multislice ptychography reconstructions \cite{chen_electron_2021, denzer2025optimizing}, while the amplitude change scales with $\mathrm{Z}^{\sim 1.94}$, similar to high-angle annular darkfield \cite{hartel1996conditions}.

We can find the amplitude contrast transfer function (ACTF) using $F_{\mathrm{a}}(\boldsymbol{\omega})=-F_{\mathrm{a}}^*(-\boldsymbol{\omega})$,
\begin{equation}
    C_{\mathrm{WAOA}}(\boldsymbol{\rho}, \boldsymbol{\Theta})=\frac{2}{\lambda \Omega_0} A(\boldsymbol{\Theta}) \operatorname{Re} \iint A(\boldsymbol{\omega}+\boldsymbol{\Theta}) \mathrm{e}^{\mathrm{i}[\chi(\boldsymbol{\Theta})-\chi(\bm{\theta})+k \boldsymbol{\omega} \cdot \boldsymbol{\rho}]} F_{\mathrm{a}}(\boldsymbol{\omega}) \mathrm{d}^2 \boldsymbol{\omega},
\end{equation}
\begin{equation}
    \tilde{C}_{\mathrm{WAOA}}(\boldsymbol{\omega}, \boldsymbol{\Theta})=\frac{\operatorname{ACTF}(\boldsymbol{\omega}, \boldsymbol{\Theta}) F_{\mathrm{a}}(\boldsymbol{\omega})}{\lambda} ,
\end{equation}
\begin{equation}
\resizebox{\linewidth}{!}{$
\begin{gathered}
\operatorname{ACTF}(\boldsymbol{\omega}, \boldsymbol{\Theta})=\frac{1}{2 \Omega_0} A(\boldsymbol{\Theta})\left\{A(\boldsymbol{\omega}-\boldsymbol{\Theta}) e^{-i[\chi(\boldsymbol{\omega}-\boldsymbol{\Theta})-\chi(-\boldsymbol{\Theta})]}+A(\boldsymbol{\omega}+\boldsymbol{\Theta}) e^{+i[\chi(\boldsymbol{\omega}+\boldsymbol{\Theta})-\chi(\boldsymbol{\Theta})]}\right\}.
\end{gathered}
$}
\end{equation}

For the special case of axial illumination ($\boldsymbol{\Theta}=0$) the ACTF simplifies to $\cos (\chi(\boldsymbol{\omega}))$ with an information cutoff at $\alpha$. For only defocus as the aberration, it can be shown that $\operatorname{ACTF}(\boldsymbol{\omega}, \boldsymbol{\Theta})$ is also conjugate symmetric about $\boldsymbol{\Theta}$, and
\begin{equation}
\resizebox{\linewidth}{!}{$
\begin{gathered}
\operatorname{ACTF}(\boldsymbol{\omega}, \boldsymbol{\Theta})\\
=\frac{1}{2 \Omega_0} A(\boldsymbol{\Theta})\left\{A(\boldsymbol{\omega}-\boldsymbol{\Theta}) e^{+1 / 2 i k_0 \Delta f |\boldsymbol{\omega}|^2}+A(\boldsymbol{\omega}+\boldsymbol{\Theta}) e^{-1 / 2 i k_0 \Delta f |\boldsymbol{\omega}|^2}\right\} e^{-i(\Delta f \bm{\theta}) \cdot\left(k_0 \boldsymbol{\omega}\right)} ,
\end{gathered}
$}
\end{equation}
\begin{equation}
    \operatorname{ACTF}(\boldsymbol{\omega},-\boldsymbol{\Theta})^*=\operatorname{ACTF}(\boldsymbol{\omega},+\boldsymbol{\Theta}).
\end{equation}

The amplitude contrast images from each detector pixel obey the same shifts as the phase contrast images and so will also contribute to the contrast of tilt-corrected imaging methods. Importantly, in a weak-phase weak-amplitude object, both contributions exist in the bright field image. The contribution to the tcBF image from amplitude contrast is
\begin{equation}
    \begin{aligned}
\operatorname{ACTF}_{\mathrm{tcBF}}(\boldsymbol{\omega}) & =\frac{1}{2} \sum_{\boldsymbol{\Theta}} \operatorname{ACTF}^{(\mathrm{tc})}(\boldsymbol{\omega},+\boldsymbol{\Theta})+\operatorname{ACTF}^{(\mathrm{tc})}(\boldsymbol{\omega},-\boldsymbol{\Theta})\\
&=\sum_{\boldsymbol{\Theta}} \Re \mathrm{e}\left(\operatorname{ACTF}^{(\mathrm{tc})}(\boldsymbol{\omega}, \boldsymbol{\Theta})\right) \\
& =\mathbb{A}(\boldsymbol{\omega}) \cos \left(1 / 2 k_0 \Delta f |\boldsymbol{\omega}|^2\right).
\end{aligned}
\end{equation}

The ACTF of tcBF shares the same envelope function as the phase contrast contribution but follows a cosine modulation instead of a sine. This implies a weak-phase weak amplitude object will still produce contrast at zero defocus from the amplitude contribution, and the average (zero frequency) contrast will be positive (dark-atom, due to the sign convention of Equation \ref{contrast_def} regardless of the defocus. Recall that the PCTF will change sign when the defocus changes sign.

\begin{figure*}
\centering
\includegraphics[width=\linewidth]{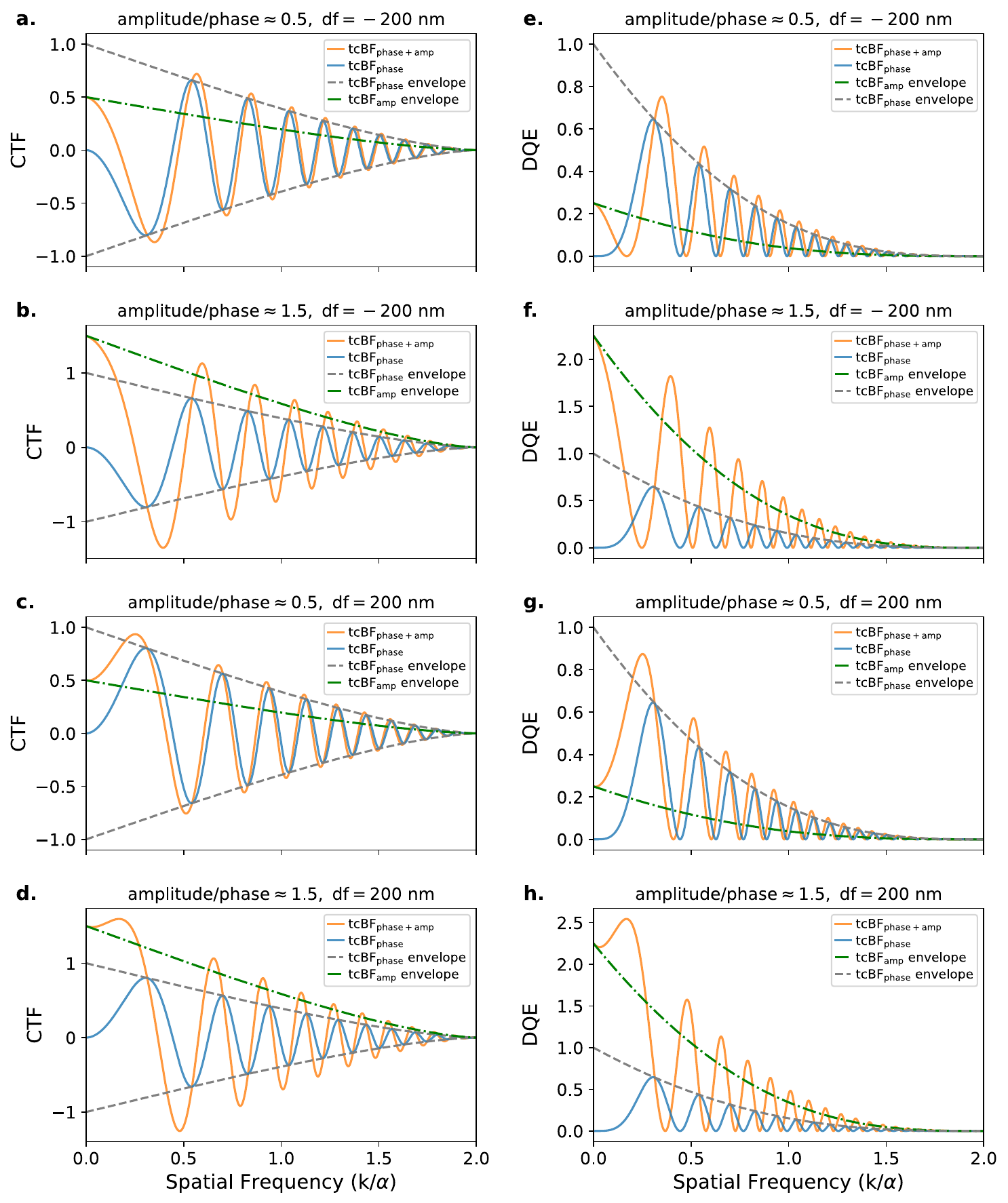}
\caption{CTF (a-d) and DQE (e-h) of tcBF under the weak phase, weak amplitude object approximation at 300 keV and 10 mrad convergence semi-angle. Amplitude/phase ratio is assumed to be 0.5 and 1.5, and defocus is assumed to be -200 nm and 200 nm respectively. DQE here is relative to the PCTF case, so can exceed 1 when the amplitude channel is added.} 
\label{fig: info_fig7_actf}
\end{figure*}

Both phase contrast and amplitude contrast are simultaneously present in the bright-field image, and because the PCTF and ACTF have the same symmetry with respect to opposite detector positions there is no strategy for separating these two contributions from a single dataset. In Figure \ref{fig: info_fig7_actf} we show the total CTFs for objects with different ratios of phase to amplitude scattering factor for both over- and under-focus conditions. The amount of contrast enhancement or reduction depends on the relative magnitude of the amplitude contribution $v s$ the phase contribution. Here, we have assumed that the anti-Friedel scattering factor follows the same shape as the symmetric component for simplicity. Such an assumption was common in early electron diffraction literature \cite{hashimoto1962anomalous}, though more accurate absorptive form factors are now available \cite{weickenmeier1991computation}. The sign of the defocus impacts which spatial frequencies have constructive or destructive combinations of amplitude and phase contrast. This can be especially beneficial when the sample contains both light and heavy elements because of the different Z-dependence, akin to the use of negative-$C_s$ imaging to ensure the phase and amplitude contributions are both positive \cite{jia2010benefit}. We note that that since the ACTF maximizes at zero defocus, for samples of thickness greater than the mean free path, it might be more beneficial to choose in-focus imaging as opposed to defocused imaging.

\begin{figure*}
\centering
\includegraphics[width=0.8\linewidth]{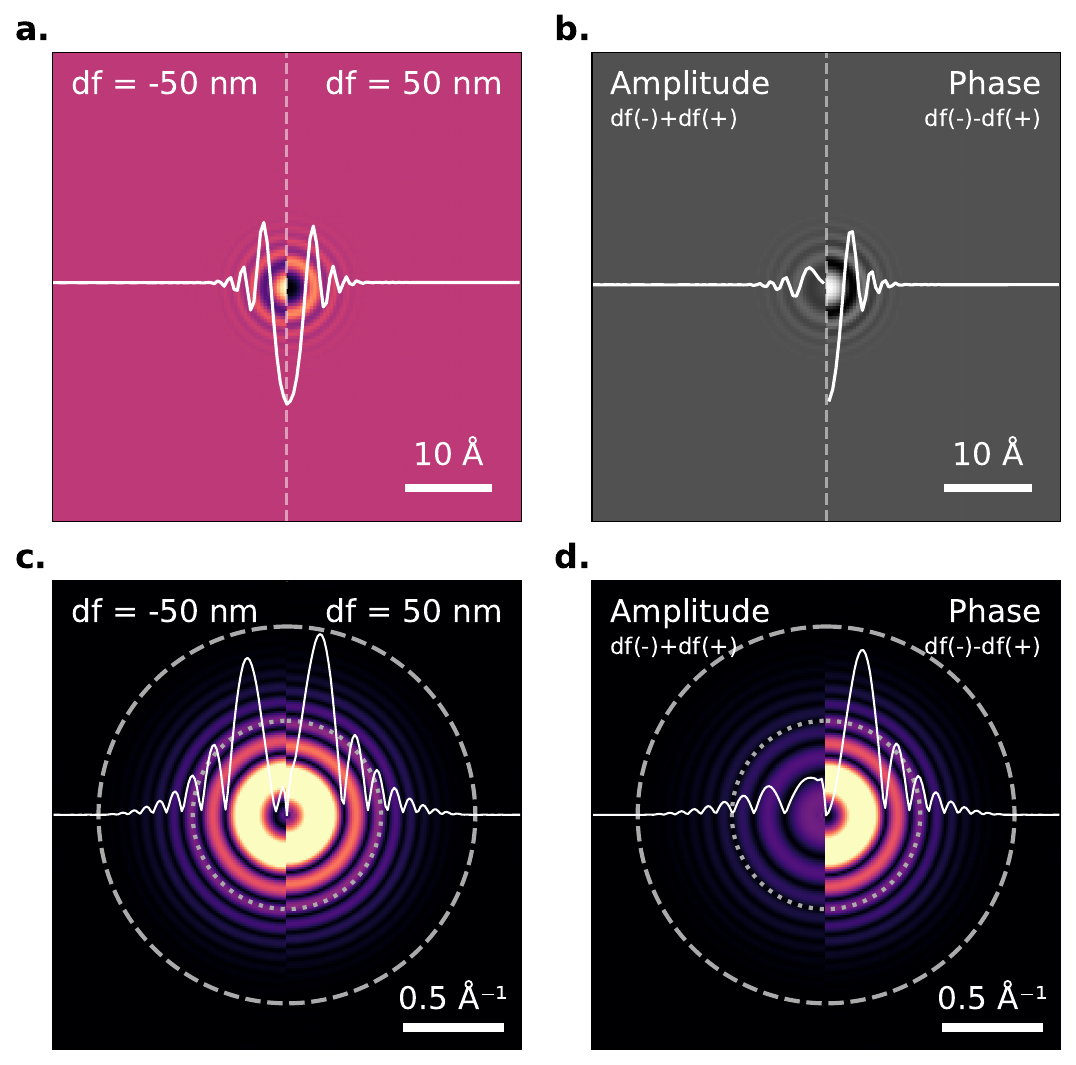}
\caption{Simulated phase and amplitude contrast from a single U atom imaged at 300 keV and 30 mrad convergence semi-angle. (a) tcBF at exactly opposite defocus values 50 and -50 nm , with noticeable mismatch due to the amplitude contrast as seen in the Fourier transform. The line profile of $\mathrm{df}=50 \mathrm{~nm}$ has been inverted to improve visualization. (c). The phase and amplitude components can be extracted by summation and subtraction of the over and under focus data as shown in (b) and (d).} 
\label{fig: info_fig8_singleU}
\end{figure*}

\begin{figure*}
\centering
\includegraphics[width=0.8\linewidth]{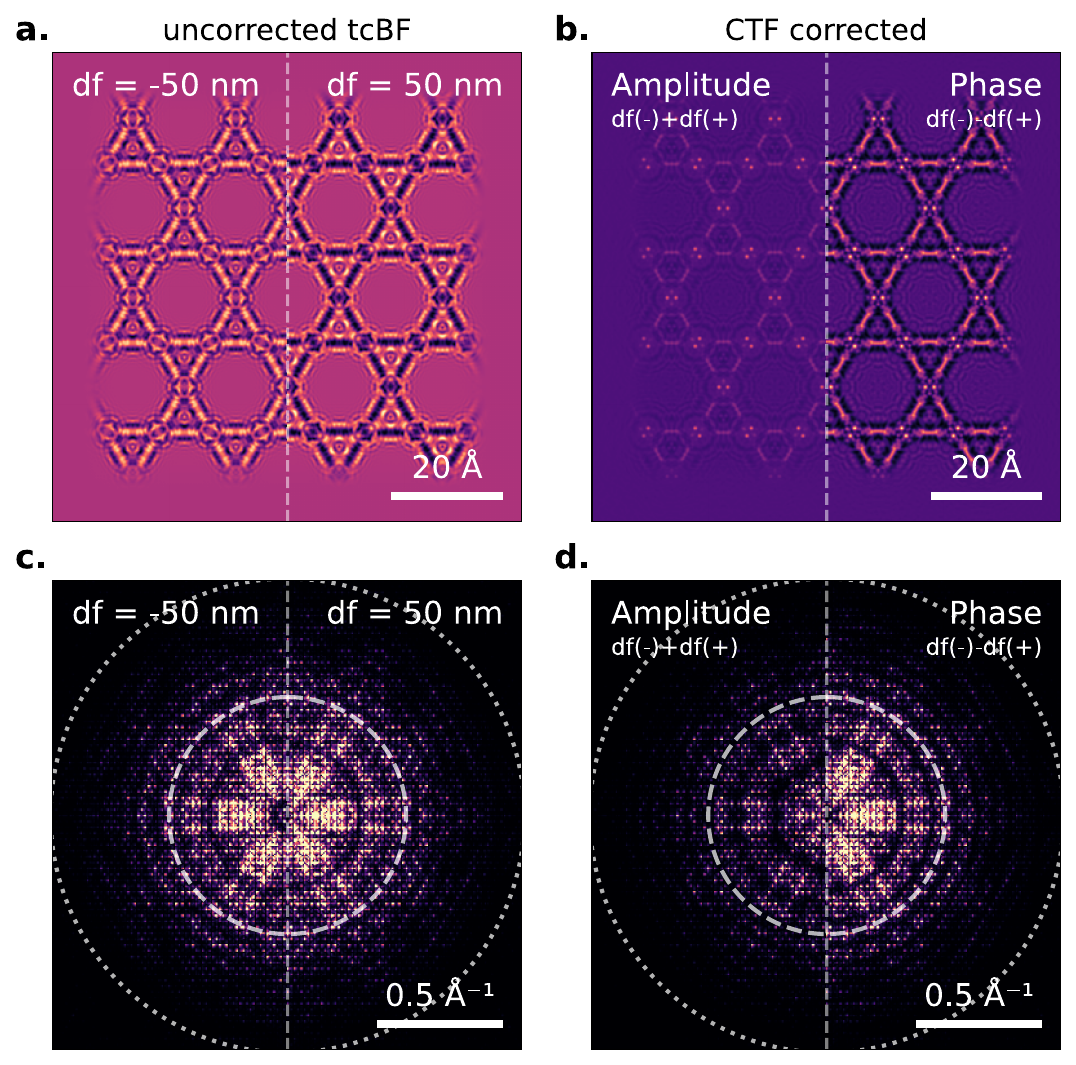}
\caption{Phase and Amplitude contrast of a simulated 4-nm-thick MOF sample potential imaged at 300 keV and 30 mrad convergence semi-angle. (a)(c) tcBF at exactly opposite defocus values 50 and -50 nm . (b)(d) The extracted phase and amplitude components can be CTF corrected accordingly to recover the atomic structure. Note that while the overall amplitude contrast has weaker intensity, the relative brightness of Zr atoms is higher because of the higher Z -dependence.} 
\label{fig: info_fig9_MOF}
\end{figure*}

Separation of the amplitude and phase contributions is possible in principle by comparing datasets taken with exactly opposite sign of defocus. Summation of the under- and over-focused reconstructions will cancel the phase-contrast contributions and leave only the amplitude, while subtraction will extract the phase contrast. We demonstrate this asymmetry of the phase and amplitude contributions with respect to the sign of defocus on simulations of a single $U$ atom and a 4-nm-thick MOF, as shown in Figure \ref{fig: info_fig8_singleU} and Figure \ref{fig: info_fig9_MOF}. Amplitude images of the MOF in Figure 9 show greater contrast between the heavy Zr atoms and the organic linker compared to the phase image. The Fourier transforms of the phase and amplitude contributions show the expected $\sin$ and $\cos$ modulation. However, acquisition of two datasets with exactly opposite defocus is extremely challenging and noise-prone in experiments, risking numerically ill-conditioned results.

The amplitude contrast of tcDPC can be similarly derived by symmetry and is found to have a sine modulation. We note that there is no amplitude contrast in tcDPC when defocus is zero (equivalent to ordinary DPC).
\begin{equation}
\begin{aligned}
\operatorname{ACTF}_{\mathrm{tcDPC}}(\boldsymbol{\omega})
   &= \frac{1}{2} \sum_{\boldsymbol{\Theta}}
      \operatorname{ACTF}^{(\mathrm{tc})}(\boldsymbol{\omega},+\boldsymbol{\Theta})
      - \operatorname{ACTF}^{(\mathrm{tc})}(\boldsymbol{\omega},-\boldsymbol{\Theta}) \\
      &= i \sum_{\boldsymbol{\Theta}}
      \Im \mathfrak{m}\!\left(\operatorname{ACTF}^{(\mathrm{tc})}(\boldsymbol{\omega}, \boldsymbol{\Theta})\right) \\
   &= i \big[\mathbb{A}(\boldsymbol{\omega})-\mathbb{A}(2 \boldsymbol{\omega})\big]
      \sin \!\left(\frac{1}{2} k_0 \Delta f |\boldsymbol{\omega}|^2\right).
\end{aligned}
\end{equation}

The tcDPC ACTF is purely imaginary, and thus produces a differential signal as in tcDPC which must be integrated to produce an amplitude image.

\subsection{Contrast from Scattering in Dark Field}
Inspecting Equation \ref{contrast_components}, only the third term, the incoherent amplitude scattering, is nonzero for  $|\boldsymbol{\Theta}|>{\alpha}$ and thus is the only contribution to the dark-field signal:
\begin{equation}
\resizebox{\linewidth}{!}{$
\begin{gathered}
C_{\mathrm{DF}}(\boldsymbol{\rho})=-\frac{1}{4 \pi^2 \Omega} \operatorname{Re} \sum_{n=0}^{\infty} \frac{k_n}{k} k_n^2 \iint A(\bm{\theta}) D(\boldsymbol{\Theta}) A\left(\bm{\theta}^{\prime}\right) \mathrm{e}^{\mathrm{i}\left[\chi_n\left(\bm{\theta}^{\prime}\right)-\chi_n(\bm{\theta})+k_n\left(\bm{\theta}-\bm{\theta}^{\prime}\right) \rho+\frac{k_n z_0\left(|\bm{\theta}|^2-|\bm{\theta}^{\prime}|^2\right)}{2}\right]} \\
\left.F_n\left(\boldsymbol{\Theta}, \bm{\theta}^{\prime}\right) F_n^*(\boldsymbol{\Theta}, \bm{\theta})\right\} \mathrm{d}^2 \bm{\theta} \mathrm{~d}^2 \bm{\theta}^{\prime} \mathrm{d}^2 \boldsymbol{\Theta}.
\end{gathered}
$}
\end{equation}

There are three notable differences from the second (amplitude contrast) term: i) the aperture functions and aberration functions only associate the two incident angles $\boldsymbol{\theta}$ and $\boldsymbol{\theta}^{\prime}$, and not the detector position $\boldsymbol{\Theta}$; ii) the spatial frequency of the contrast transfer is given by $|\boldsymbol{\Omega}|=|\boldsymbol{\theta}-\boldsymbol{\theta}^{\prime}| \in [0,2 \alpha]$; iii) the exponential phase factor contains all inelastic channels $\mathrm{n}>0$, over which we perform an incoherent summation. As a result, unlike the other two terms, for the single-pixel dark field images we cannot immediately extract a real-space shift operator that would lead to a tilt-corrected imaging method. Summation over all azimuthal angles result in the ordinary annular dark-field (ADF) imaging (see, for example, Hartel et al \cite{hartel1996conditions} for details).

Typically, the structure factor decays quickly as a function of scattering angle, so contrast at a given dark-field detector position is primarily due to the scattering from the nearest position within the bright-field disk. Thus, the image will obey the same shift as the nearest bright-field position. In other words, there is still rich angle-dependent and depth-dependent information encoded in this dark field region that can be extracted with a pixelated detector (and used in ptychography) beyond a simple incoherent summation. In the next section we show how we can obtain dark-field depth sectioning with the 4D dataset even by just shifting the azimuthal components accordingly.

\subsection{One-shot Depth Sectioning enabled by 4D-STEM}
One-shot depth sectioning for tcBF or acBF is naturally enabled by adjusting the amount of shifts $\Delta f+\mathrm{d} z$ according to the z height of interest in the shift operator $e^{-i((\Delta f+\mathrm{d} z) \Theta) \cdot\left(k_0 \omega\right)}$ as shown in Equation \ref{df_shift} during the reconstruction. This can be understood by the parallax effect \cite{chen_imaging_2024,seifer_flexible_2021}, which has been used for 3D reconstructions in the shadow montage method \cite{seiferShadowMontageConeBeam2025}. Depth sectioning for dark field, which we term tilt-corrected dark-field imaging (tcDF), can be enabled by shifting according to $\Delta f+\mathrm{d} z$ as well. Nevertheless, instead of shifting each pixel by $\boldsymbol{\Theta}$, we sum over all pixels with $|\boldsymbol{\Theta}|>{\alpha}$ and only shift azimuthally by $(2 \alpha / 3)(\Delta f+\mathrm{d} z)$. This shift amount is given by the expectation of the distance from the optical axis of a randomly chosen incident angle $\boldsymbol{\theta}^{\prime}$ within the ${\alpha}$ aperture with probability $P(r) = {(2\pi r) \mathrm{d}r}/{\pi \alpha^2}$:
\begin{equation}
    \mathbb{E}\left[\left|\boldsymbol{\theta}^{\prime}\right|\right]=\int_0^\alpha r \cdot P(r) =\int_0^\alpha r \cdot \frac{2 r}{\alpha^2} d r=\frac{2 \alpha}{3}.
\end{equation}

\begin{figure*}[h]
\centering
\includegraphics[width=\linewidth]{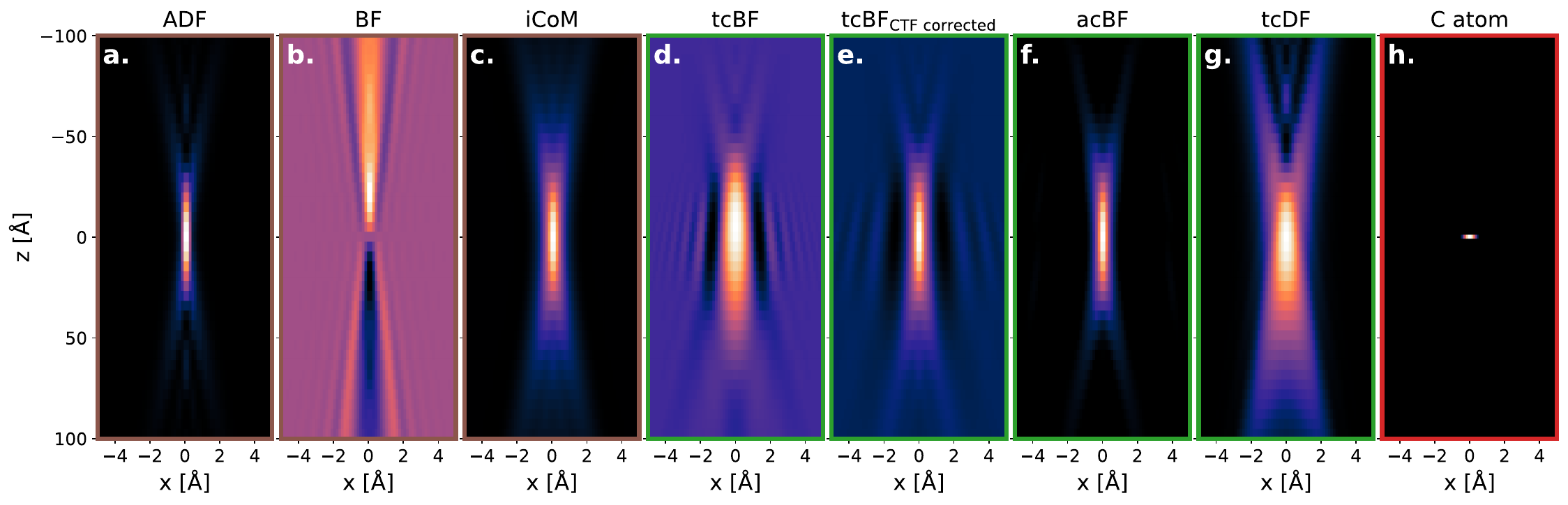}
\caption{Depth sectioning of a simulated C atom at 300 keV and 30 mrad convergence semiangle to explore the approximate 3D point spread function. (a-c) Conventional serial depth sectioning using a through-focal series from multiple $\mathrm{ADF}, \mathrm{BF}$, and iCOM images. (d-g) One-shot depth sectioning by tcBF , tcBF ( CTF corrected), acBF , and tcDF from a single 4 D dataset at a defocus of 10 nm . (h) Ground truth atomic potential of a C atom.} 
\label{fig: info_fig10_tcDF}
\end{figure*}

The physical intuition becomes obvious if we compare the form of Equation (33) to the coherent terms in Equation (7) where the spatial frequency of contrast transfer is determined by $\boldsymbol{\omega}=\boldsymbol{\theta}-\boldsymbol{\Theta}$. For each incident angle $\boldsymbol{\theta}$, only the other scattering angle decides the shift, which in the case of tcDF is $\boldsymbol{\theta}^{\prime}$. In Figure 10, we demonstrate the available depth-sectioning power of a single 4D-STEM dataset compared to conventional depth sectioning using focal series.

Again, the existence of the tcDF signal, and its depth sensitivity indicates useful information that can be measured directly, or used in iterative ptychography methods that is explicitly excluded from WPOA analyses.

\section{Discussion}
In this work, we have revisited and extended the contrast transfer formalism originally developed \cite{rose_nonstandard_1976, urban2015quest} by Rose, which identifies different classes of contrast mechanisms based on the symmetry of the detectors even in the presence of multiple scattering, adapting it to the emerging capabilities of 4D scanning transmission electron microscopy (4D-STEM). By decomposing the phase contrast transfer into symmetric and antisymmetric components under the weak phase object approximation (WPOA), we have established a comprehensive imaging framework that identifies information missed by traditional direct ptychography methods such as SSB and iDPC, and provides a general framework to recover that missing information. Keeping only the coherent phase term of equation \ref{contrast_components}, the simplest of our approximations correct for the defocus-induced image shift that blurs conventional bright field and DPC imaging, leading to tilt-corrected bright field (tcBF) and tilt-corrected differential phase contrast (tcDPC) imaging. These in turn are combined, leading to the formulation of aberration-corrected bright field (acBF) imaging, which provides enhanced and continuous phase contrast transfer up to the aperture-limited information limit of $2 \alpha$.

A key advantage of acBF lies in its ability to circumvent the information loss typically associated with zero crossings in the sine-form CTF of conventional BF imaging. By coherently integrating the symmetric (tcBF) and antisymmetric (tcDPC) components of the scattering signal, acBF recovers spatial frequency information across the full area of the bright field disk, achieving the maximal possible phase contrast transfer for a given acquisition condition under the WPOA. AcBF offers a phase imaging modality that is largely insensitive to defocus-induced contrast modulations or even reversals. This approach is particularly beneficial for imaging weakly scattering materials, such as light-element systems and beam-sensitive biological specimens, where the signal to noise ratio (SNR) is low due to low dose and iterative reconstruction methods, e.g., ptychography can fail to converge.

The tilt-corrected imaging methods thus emerge as a new family of direct ptychography. As an initial step, tilt correction recovers coherent contrast transfer by unfolding the entangled symmetric and antisymmetric components in the complex CTF and identifies their presence in different interference regimes, i.e., the triple overlap and double overlap regions in detector space. Consequently, separate or combined usage of these channels can be designed at will to produce different phase retrieval methods across varied aperture geometries and coherence conditions.

Beyond the weak phase object approximation, we have also analyzed the contribution due to the amplitude change of the scattered wave. This scattering arises from the interference between elastic and inelastic (or plural) scattering components and can have a nonlinear dependence on the sample potential and hence greater Z-sensitivity. These contributions are small for light elements and very thin samples but are noticeable even in single atom simulations for $\mathrm{Z}>6$. As has been previously noted for conventional TEM imaging \cite{kirkland_advanced_2020}, they cannot (or at least should not) be neglected in quantitative analyses.

Our analysis shows that this contribution to the information transfer is symmetric with respect to the sign of defocus, implying that an over-focused probe should be preferred so that their interference can be constructively harnessed to enhance contrast at select spatial frequencies, as in CTEM. This is coincidentally in line with the typical practice in multislice electron ptychography, where most reported works have used an overfocus condition in order to avoid focal crossover occurring within the object.

The ACTF provides a foundation for analyzing intensity-modifying effects in materials with significant scattering potential, while also maintaining compatibility with phase-sensitive imaging schemes. In particular, the coherent amplitude contributions may be exploited in hybrid imaging strategies that integrate both amplitude and phase information. Iterative ptychography methods model the object as modifying both the amplitude and phase of the wave and so can recover both contributions. However, our results indicate that these two components are difficult to disentangle from a single scan at a single defocus value, and the choice of experimental parameters affects how strongly each component is present in the 4D-STEM data.

A more complete description of the contrast from thick samples, especially stained biological sections, would require the use of the thick weak phase/amplitude model \cite{seki_theoretical_2018,seki2022linear}, wherein the PCTF and ACTF derived above is integrated over a range of defocus values to approximate the variation of the response as the probe transits the sample. For imaging at atomic resolution the defocus spread of the probe due to a finite energy spread of the source and chromatic aberration of the optics may become an important factor limiting resolution, and this may also be accounted for by a weighted average of the CTFs over the defocus spread. While multislice ptychography can recover 3D information by deconvolving the appropriate CTF at each layer, our results still establish limits on how strongly such information is encoded by the physics of scattering into the measured data. Some information transfer beyond what is indicated by our analysis may yet encode such information, particularly dynamical scattering effects which we exclude by taking the first Born approximation.

In the context of incoherent scattering, we have introduced the concept of tilt-corrected dark field (tcDF) imaging. While incoherent contrast lacks the phase information present in the coherent channel, we demonstrate that tcDF still enables extraction of further information by correcting the azimuth-dependent shifts of the darkfield scattering. tcDF provides additional information for one-shot depth sectioning of thick and heterogeneous specimens. This points to a broader generalization of the 4D-STEM contrast formalism, wherein coherent and incoherent mechanisms are not mutually exclusive but rather complementary under appropriate data processing strategies.

Despite these advances, limitations remain. The current simple algorithms neglect multiple scattering in the sample and assume idealized coherence, partial temporal, or spatial coherence effects, which may become significant under practical conditions. Partial coherence effects will become significant when the detector segments are no longer very small compared to the probeforming semi-angle, such as in segmented detectors with fewer than 100 segments. Additionally, while the theoretical formulations are general, numerical implementation and signal-to-noise limitations in experimental data may hinder full realization of the proposed imaging modes. Future work should aim to integrate these theoretical tools with regularized reconstruction methods and denoising techniques, to fully exploit the potential of acBF and related methods in real-world microscopy applications.

\section{Conclusion}
In this paper, we revisit the origin of contrast transfer mechanisms from a quantum theory of electron scattering, and explore where information is encoded in 4D-STEM data sets. The theory naturally divides the general scattering problem into coherent phase, coherent amplitude and incoherent amplitude terms. By decomposing the coherent phase signals into symmetric and antisymmetric components under the weak phase object approximation, we formulate aberrationcorrected bright-field (acBF) imaging as a physically grounded, analytically tractable method that unifies and extends tcBF and DPC imaging. AcBF maximizes the recoverable phase information within the bright-field disk, achieving continuous and interpretable contrast transfer across the full $2 \alpha$ support, while remaining robust to aberrations and defocus modulations. AcBF can serve as a crude but fast approximation to ptychography in the case of weak phase objects and low-dose imaging conditions.

We further expanded the framework of contrast beyond linear phase contrast transfer mechanisms by identifying a coherent amplitude contrast transfer function (ACTF) derived from the generalized optical theorem. This component-often neglected in standard imaging model scan be constructively utilized under specific probe conditions to enhance contrast in strongly scattering or thick samples. In addition, we introduce tilt-corrected dark-field (tcDF) imaging as a practical extension for incoherent contrast, providing access to depth-sensitive information through azimuthal tilt alignment within only one 4D dataset. The presence of useful information in these two scattering terms bridges the gap between the weak-phase direct ptychography methods and full-field iterative ptychography that can exploit this additional contrast for more dose-efficient recovery of information from the sample.

This framework generalizes phase contrast theory in conventional/scanning transmission electron microscopy to 4D-STEM and provides analytical models and insights into novel quantitative imaging methods of varied aperture and detector geometries under different noise characteristics, sample thicknesses and coherence conditions, e.g., full-field iterative ptychography, which blindly exploits all above contrast mechanisms. Taken together, this work provides a unified and physically-grounded framework for understanding and optimizing contrast in 4D-STEM. By systematically separating and recombining the various scattering channels -- symmetric/antisymmetric, coherent/incoherent, and phase/amplitude -- we developed a computationally-tractable platform of direct algorithms suitable for modern detector architectures and aberration-corrected probes.

\appendix
\section{Detailed derivations for tcBF, tcDPC and acBF}\label{appendix_part3_A}
\setcounter{figure}{0} % reset figure numbering
\renewcommand{\thefigure}{A\arabic{figure}} % ensure format is A1, A2, etc.

We start from constructing the envelope function for the overlap disks. We define the normalized area of the overlap of two disks separated by $\omega=|\boldsymbol{\omega}|$ to be $\mathbb{A}(\boldsymbol{\omega})$, given by geometry,
\begin{equation}
    \mathbb{A}(\boldsymbol{\omega})=\frac{2}{\pi}\left[\cos ^{-1}\left(\frac{\omega}{2}\right)-\frac{\omega}{2} * \sqrt{1-\left(\frac{\omega}{2}\right)^2}\right], \quad \omega<2 \alpha.
\end{equation}

This is the envelope for ADF and tcBF. For the overlap of 2 disks when 3 disks are present, i.e., the trotters, the envelope for DPC or SSB is,
\begin{equation}
    \mathbb{B}(\boldsymbol{\omega})=\mathbb{A}(\boldsymbol{\omega})-\mathbb{A}(2 \boldsymbol{\omega}).
\end{equation}

\begin{figure}
    \centering
    \includegraphics[width=\linewidth]{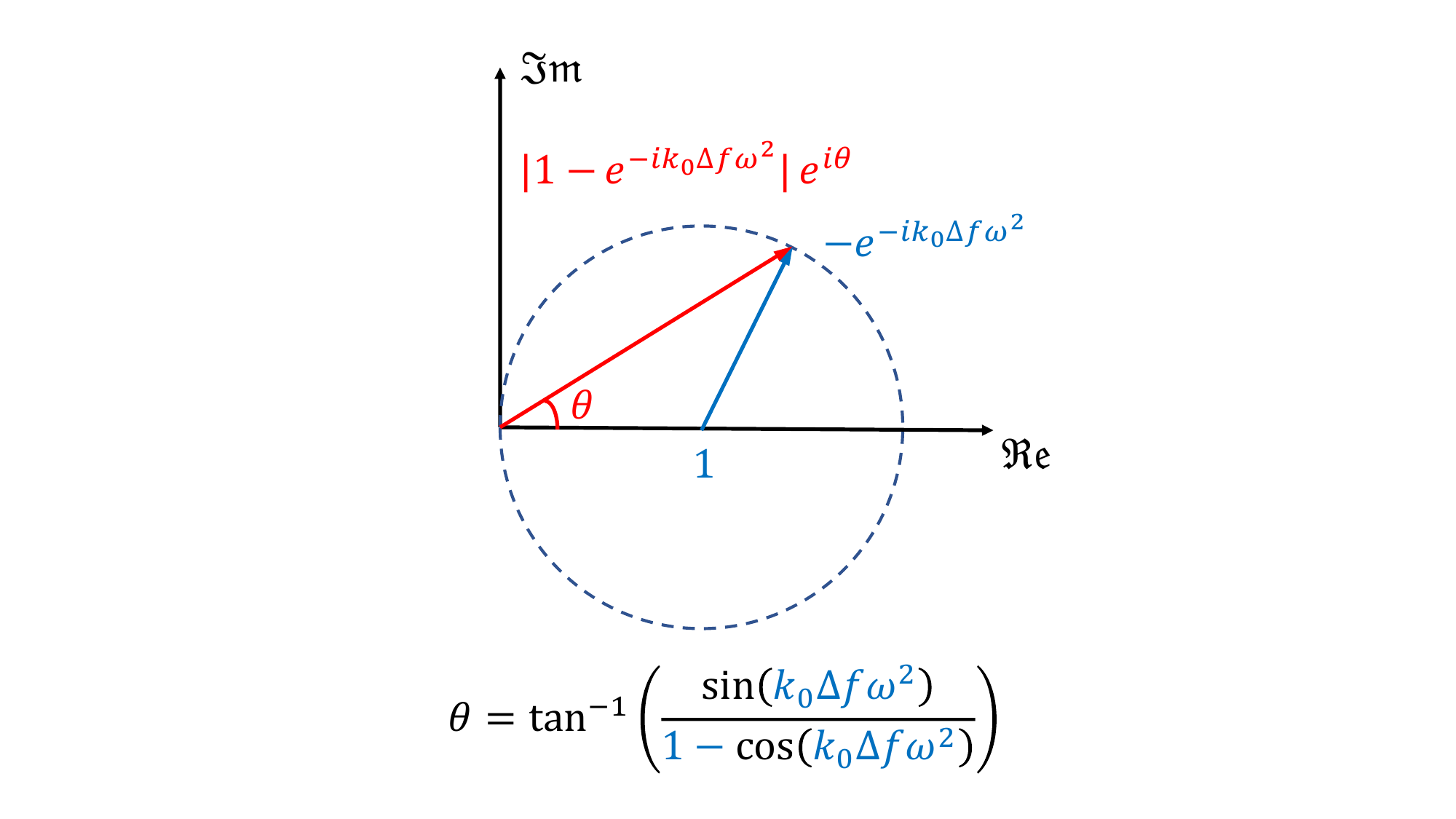}
    \caption{Maximal phase contrast transfer is achieved when the whole modulus of the complex CTF is transferred as shown in Equation \ref{equ:acbf_do1}, \ref{equ:acbf_do11} ,\ref{equ:acbf_to} ,\ref{equ:acbf_all}.}
    \label{fig:info_A1}
\end{figure}

\subsection{tcBF}
\begin{equation}
\resizebox{\linewidth}{!}{%
$\begin{aligned}
\operatorname{PCTF}_{\mathrm{tcBF}}(\boldsymbol{\omega}) 
&=  \frac{1}{2} \sum_{\boldsymbol{\Theta}} 
   \operatorname{PCTF}^{(\mathrm{tc})}(\boldsymbol{\omega},+\boldsymbol{\Theta})
   +\operatorname{PCTF}^{(\mathrm{tc})}(\boldsymbol{\omega},-\boldsymbol{\Theta}) \\
&= \sum_{\boldsymbol{\Theta}} 
   \Re \mathrm{e}\!\left(\operatorname{PCTF}^{(\mathrm{tc})}(\boldsymbol{\omega}, \boldsymbol{\Theta})\right) \\
&= \Re \mathrm{e}\!\left[\sum_{\boldsymbol{\Theta}}
   \Big(\frac{i}{\Omega_0} A(\boldsymbol{\Theta})
   \{A(\boldsymbol{\omega}-\boldsymbol{\Theta}) e^{+\frac{i}{2} k_0 \Delta f |\boldsymbol{\omega}|^2}
   - A(\boldsymbol{\omega}+\boldsymbol{\Theta}) e^{-\frac{i}{2} k_0 \Delta f |\boldsymbol{\omega}|^2}\}\Big)\right] \\
&= \Re \mathrm{e}\!\left[\frac{i}{\pi \alpha^2} 
   \int_0^\alpha 2 \pi \Theta\, d \Theta\, A(\boldsymbol{\Theta})
   \{A(\boldsymbol{\omega}-\boldsymbol{\Theta}) e^{+\frac{i}{2} k_0 \Delta f |\boldsymbol{\omega}|^2}
   - A(\boldsymbol{\omega}+\boldsymbol{\Theta}) e^{-\frac{i}{2} k_0 \Delta f |\boldsymbol{\omega}|^2}\}\right] \\
&= -\frac{1}{\pi \alpha^2} \int_0^\alpha 2 \pi \Theta\, d \Theta\, 
   A(\boldsymbol{\Theta})
   \{A(\boldsymbol{\omega}-\boldsymbol{\Theta})+A(\boldsymbol{\omega}+\boldsymbol{\Theta})\}
   \sin \!\left(\frac{1}{2} k_0 \Delta f |\boldsymbol{\omega}|^2\right) \\
&= -\mathbb{A}(\boldsymbol{\omega})\, \sin \!\left(\frac{1}{2} k_0 \Delta f |\boldsymbol{\omega}|^2\right).
\end{aligned}$%
}
\end{equation}

\begin{equation}
\resizebox{\linewidth}{!}{%
$\begin{aligned}
\operatorname{ACTF}_{\mathrm{tcBF}}(\boldsymbol{\omega})  =&\frac{1}{2} \sum_{\boldsymbol{\Theta}} \operatorname{ACTF}^{(\mathrm{tc})}(\boldsymbol{\omega},+\boldsymbol{\Theta})+\operatorname{ACTF}^{(\mathrm{tc})}(\boldsymbol{\omega},-\boldsymbol{\Theta})\\
=&\sum_{\boldsymbol{\Theta}} \Re \mathrm{e}\left(\operatorname{ACTF}^{(\mathrm{tc})}(\boldsymbol{\omega}, \boldsymbol{\Theta})\right) \\
= & \Re \mathrm{e}\left[\sum_{\boldsymbol{\Theta}}\left(1 / \Omega_0 A(\boldsymbol{\Theta})\left\{A(\boldsymbol{\omega}-\boldsymbol{\Theta}) e^{+1 / 2 i k_0 \Delta f |\boldsymbol{\omega}|^2}+A(\boldsymbol{\omega}+\boldsymbol{\Theta}) e^{-1 / 2 i k_0 \Delta f |\boldsymbol{\omega}|^2}\right\}\right)\right] \\
= & \Re \mathrm{e}\left[\frac{1}{\pi \alpha^2} \int_0^\alpha 2 \pi \Theta d \Theta A(\boldsymbol{\Theta})\left\{A(\boldsymbol{\omega}-\boldsymbol{\Theta}) e^{+1 / 2 i k_0 \Delta f |\boldsymbol{\omega}|^2}+A(\boldsymbol{\omega}+\boldsymbol{\Theta}) e^{-1 / 2 i k_0 \Delta f |\boldsymbol{\omega}|^2}\right\}\right] \\
= & {\left[\frac{1}{\pi \alpha^2} \int_0^\alpha 2 \pi \Theta d \Theta A(\boldsymbol{\Theta})\{A(\boldsymbol{\omega}-\boldsymbol{\Theta})+A(\boldsymbol{\omega}+\boldsymbol{\Theta})\} \cos \left(1 / 2 k_0 \Delta f |\boldsymbol{\omega}|^2\right)\right] } \\
= & \mathbb{A}(\boldsymbol{\omega}) \cos \left(1 / 2 k_0 \Delta f |\boldsymbol{\omega}|^2\right).
\end{aligned}$%
}
\end{equation}

\subsection{tcDPC}
\begin{equation}
\resizebox{\linewidth}{!}{%
$\begin{aligned}
\operatorname{PCTF}_{\mathrm{t c D P C}} (\boldsymbol{\omega})= &\frac{1}{2} \sum_{\boldsymbol{\Theta}} \operatorname{PCTF}^{(\mathrm{tc})}(\boldsymbol{\omega},+\boldsymbol{\Theta})-\operatorname{PCTF}^{(\mathrm{tc})}(\boldsymbol{\omega},-\boldsymbol{\Theta})\\
=&i \sum_{\boldsymbol{\Theta}} \Im \mathfrak{m}\left(\operatorname{PCTF}^{(\mathrm{tc})}(\boldsymbol{\omega}, \boldsymbol{\Theta})\right) \\
= & i\left(\Im \mathfrak{m}\left[\sum_{\boldsymbol{\Theta}}\left(i / \Omega_0 A(\boldsymbol{\Theta})\left\{A(\boldsymbol{\omega}-\boldsymbol{\Theta}) e^{+1 / 2 i k_0 \Delta f |\boldsymbol{\omega}|^2}-A(\boldsymbol{\omega}+\boldsymbol{\Theta}) e^{-1 / 2 i k_0 \Delta f |\boldsymbol{\omega}|^2}\right\}\right)\right]\right) \\
= & i\left(\Im \mathfrak{m}\left[\frac{i}{\pi \alpha^2} \int_0^\alpha 2 \pi \Theta d \Theta A(\boldsymbol{\Theta})\left\{A(\boldsymbol{\omega}-\boldsymbol{\Theta}) e^{+1 / 2 i k_0 \Delta f |\boldsymbol{\omega}|^2}-A(\boldsymbol{\omega}+\boldsymbol{\Theta}) e^{-1 / 2 i k_0 \Delta f |\boldsymbol{\omega}|^2}\right\}\right]\right) \\
= & i\left[\frac{1}{\pi \alpha^2} \int_0^\alpha 2 \pi \Theta d \Theta A(\boldsymbol{\Theta})\{A(\boldsymbol{\omega}-\boldsymbol{\Theta})-A(\boldsymbol{\omega}+\boldsymbol{\Theta})\} \cos \left(1 / 2 k_0 \Delta f |\boldsymbol{\omega}|^2\right)\right] \\
= & i[\mathbb{A}(\boldsymbol{\omega})-\mathbb{A}(2 \boldsymbol{\omega})] \cos \left(1 / 2 k_0 \Delta f |\boldsymbol{\omega}|^2\right).
\end{aligned}$%
}
\end{equation}

\begin{equation}
\resizebox{\linewidth}{!}{%
$\begin{aligned}
\operatorname{ACTF}_{\mathrm{t c D P C}}  (\boldsymbol{\omega})=&\frac{1}{2} \sum_{\boldsymbol{\Theta}} \operatorname{ACTF}^{(\mathrm{tc})}(\boldsymbol{\omega},+\boldsymbol{\Theta})-\operatorname{ACTF}^{(\mathrm{tc})}(\boldsymbol{\omega},-\boldsymbol{\Theta})\\
=&i \sum_{\boldsymbol{\Theta}} \Im \mathfrak{m}\left(\operatorname{ACTF}^{(\mathrm{tc})}(\boldsymbol{\omega}, \boldsymbol{\Theta})\right)  \\
= & i\left(\Im \mathfrak{m}\left[\sum_{\boldsymbol{\Theta}}\left(1 / \Omega_0 A(\boldsymbol{\Theta})\left\{A(\boldsymbol{\omega}-\boldsymbol{\Theta}) e^{+1 / 2 i k_0 \Delta f |\boldsymbol{\omega}|^2}+A(\boldsymbol{\omega}+\boldsymbol{\Theta}) e^{-1 / 2 i k_0 \Delta f |\boldsymbol{\omega}|^2}\right\}\right)\right]\right) \\
= & i\left(\Im \mathfrak{m}\left[\frac{1}{\pi \alpha^2} \int_0^\alpha 2 \pi \Theta d \Theta A(\boldsymbol{\Theta})\left\{A(\boldsymbol{\omega}-\boldsymbol{\Theta}) e^{+1 / 2 i k_0 \Delta f |\boldsymbol{\omega}|^2}+A(\boldsymbol{\omega}+\boldsymbol{\Theta}) e^{-1 / 2 i k_0 \Delta f |\boldsymbol{\omega}|^2}\right\}\right]\right) \\
= & i\left[\frac{1}{\pi \alpha^2} \int_0^\alpha 2 \pi \Theta d \Theta A(\boldsymbol{\Theta})\{A(\boldsymbol{\omega}-\boldsymbol{\Theta})-A(\boldsymbol{\omega}+\boldsymbol{\Theta})\} \sin \left(1 / 2 k_0 \Delta f |\boldsymbol{\omega}|^2\right)\right]  \\
= & i[\mathbb{A}(\boldsymbol{\omega})-\mathbb{A}(2 \boldsymbol{\omega})] \sin \left(1 / 2 k_0 \Delta f |\boldsymbol{\omega}|^2\right).
\end{aligned}$%
}
\end{equation}

\subsection{acBF}
For acBF, we apply phase correction according to each region in frequency space.

\textbullet\ If $|\boldsymbol{\omega}+\boldsymbol{\Theta}|>\alpha$ and $|\boldsymbol{\omega}-\boldsymbol{\Theta}|<\alpha$, apply complex phase shift $e^{-1 / 2 i k_0 \Delta f |\boldsymbol{\omega}|^2}$,
\begin{equation}
    \begin{aligned}
& A(\boldsymbol{\Theta})\left\{A(\boldsymbol{\omega}-\boldsymbol{\Theta}) e^{+1 / 2 i k_0 \Delta f |\boldsymbol{\omega}|^2}-A(\boldsymbol{\omega}+\boldsymbol{\Theta}) e^{-1 / 2 i k_0 \Delta f |\boldsymbol{\omega}|^2}\right\} e^{-1 / 2 i k_0 \Delta f |\boldsymbol{\omega}|^2}  \\
& =A(\boldsymbol{\Theta})\left\{A(\boldsymbol{\omega}-\boldsymbol{\Theta})-A(\boldsymbol{\omega}+\boldsymbol{\Theta}) e^{+i k_0 \Delta f |\boldsymbol{\omega}|^2}\right\} \\
& =1.
\end{aligned}
\end{equation}

\textbullet\ If $|\boldsymbol{\omega}+\boldsymbol{\Theta}|<\alpha$ and $|\boldsymbol{\omega}-\boldsymbol{\Theta}|>\alpha$, apply complex phase shift $-e^{+1 / 2 i k_0 \Delta f |\boldsymbol{\omega}|^2}$,
\begin{equation}
    \begin{aligned}
& A(\boldsymbol{\Theta})\left\{A(\boldsymbol{\omega}-\boldsymbol{\Theta}) e^{+1 / 2 i k_0 \Delta f |\boldsymbol{\omega}|^2}-A(\boldsymbol{\omega}+\boldsymbol{\Theta}) e^{-1 / 2 i k_0 \Delta f |\boldsymbol{\omega}|^2}\right\}\left(-e^{+1 / 2 i k_0 \Delta f |\boldsymbol{\omega}|^2}\right) \\
& =A(\boldsymbol{\Theta})\left\{-A(\boldsymbol{\omega}-\boldsymbol{\Theta}) e^{i k_0 \Delta f |\boldsymbol{\omega}|^2}+A(\boldsymbol{\omega}+\boldsymbol{\Theta})\right\} \\
& =1.
\end{aligned}
\end{equation}

\textbullet\ If $|\boldsymbol{\omega}+\boldsymbol{\Theta}|<\alpha$ and $|\boldsymbol{\omega}-\boldsymbol{\Theta}|<\alpha$, only sign flipping is needed as the PCTF is already real in this region,
\begin{equation}
    \begin{aligned}
& \left|A(\boldsymbol{\Theta})\left\{A(\boldsymbol{\omega}-\boldsymbol{\Theta}) e^{+1 / 2 i k_0 \Delta f |\boldsymbol{\omega}|^2}-A(\boldsymbol{\omega}+\boldsymbol{\Theta}) e^{-1 / 2 i k_0 \Delta f |\boldsymbol{\omega}|^2}\right\}\right| \\
& \quad=\left|e^{+1 / 2 i k_0 \Delta f |\boldsymbol{\omega}|^2}-e^{-1 / 2 i k_0 \Delta f |\boldsymbol{\omega}|^2}\right| \\
& \quad=\left|\cos \left(k_0 \Delta f |\boldsymbol{\omega}|^2 / 2\right)+i \sin \left(k_0 \Delta f |\boldsymbol{\omega}|^2 / 2\right)-\cos \left(k_0 \Delta f |\boldsymbol{\omega}|^2 / 2\right)+i \sin \left(k_0 \Delta f |\boldsymbol{\omega}|^2 / 2\right)\right|  \\
& \quad=2\left|\sin \left(k_0 \Delta f |\boldsymbol{\omega}|^2 / 2\right)\right|.
\end{aligned}
\end{equation}

In practice, instead of taking the absolute value, it is numerically beneficial to apply a phase shift: $e^{-1 / 2 i k_0 \Delta f |\boldsymbol{\omega}|^2} e^{-i \tan ^{-1}\left(\frac{\sin \left(k_0 \Delta f |\boldsymbol{\omega}|^2\right)}{1+\cos \left(k_0 \Delta f |\boldsymbol{\omega}|^2\right)}\right)}$ as shown in Figure A1,
\begin{equation}
\resizebox{\linewidth}{!}{%
$\begin{aligned}
& A(\boldsymbol{\Theta})\left\{A(\boldsymbol{\omega}-\boldsymbol{\Theta}) e^{+1 / 2 i k_0 \Delta f |\boldsymbol{\omega}|^2}-A(\boldsymbol{\omega}+\boldsymbol{\Theta}) e^{-1 / 2 i k_0 \Delta f |\boldsymbol{\omega}|^2}\right\} e^{-1 / 2 i k_0 \Delta f |\boldsymbol{\omega}|^2} e^{-i \tan ^{-1}\left(\frac{\sin \left(k_0 \Delta f |\boldsymbol{\omega}|^2\right)}{1+\cos \left(k_0 \Delta f |\boldsymbol{\omega}|^2\right)}\right)}  \\
& =\mid 1-e^{-i k_0 \Delta f |\boldsymbol{\omega}|^2 \mid } \\
& =1-\cos \left(k_0 \Delta f |\boldsymbol{\omega}|^2\right)+i \sin \left(k_0 \Delta f |\boldsymbol{\omega}|^2\right)  \\
& =\sqrt{\left(1-\cos \left(k_0 \Delta f |\boldsymbol{\omega}|^2\right)\right)^2+\sin ^2\left(k_0 \Delta f |\boldsymbol{\omega}|^2\right)}  \\
& =\sqrt{2\left(1-\cos \left(k_0 \Delta f |\boldsymbol{\omega}|^2\right)\right)}  \\
& =2 \sqrt{\sin ^2\left(k_0 \Delta f |\boldsymbol{\omega}|^2 / 2\right)}.
\end{aligned}$%
}
\end{equation}

Summing the corrected transfer functions over all $\bm\Theta$ gives the full PCTF for acBF,
\begin{equation}
    \operatorname{PCTF}_{\mathrm{acBF}}(\boldsymbol{\omega})=[\mathbb{A}(\boldsymbol{\omega})-\mathbb{A}(2 \boldsymbol{\omega})]+\mathbb{A}(2 \boldsymbol{\omega})\left|\sin \left(k_0 \Delta f |\boldsymbol{\omega}|^2 / 2\right)\right|.
\end{equation}

\section{Alternative approach to acBF: Forming a complex image from tcBF and tcDPC}\label{appendix_part3_B}
\setcounter{figure}{0} % reset figure numbering
\renewcommand{\thefigure}{B\arabic{figure}} % ensure format is A1, A2, etc.

Recall that the PCTFs for tcBF and tcDPC signals at defocus $\Delta f$ are given by:

For tcBF,
\begin{equation}
    \operatorname{PCTF}_{\mathrm{tcBF}}(\boldsymbol{\omega})=-\mathbb{A}(\boldsymbol{\omega}) \sin \left(1 / 2 k_0 \Delta f |\boldsymbol{\omega}|^2\right).
\end{equation}

For tcDPC,
\begin{equation}
    \operatorname{PCTF}_{\mathrm{tcDPC}}(\boldsymbol{\omega})=i \mathbb{B}(\boldsymbol{\omega}) \cos \left(1 / 2 k_0 \Delta f |\boldsymbol{\omega}|^2\right).
\end{equation}
where for convenience we write $\mathbb{B}(\boldsymbol{\omega})=\mathbb{A}(\boldsymbol{\omega})-\mathbb{A}(2 \boldsymbol{\omega})$
Notice the envelope function $\mathbb{A}(\boldsymbol{\omega})$ and $\mathbb{B}(\boldsymbol{\omega})$ are determined only by the aperture size $\alpha$ and not the defocus. This suggests a simple form for CTF correction is possible where we do not have to measure the defocus, but instead by correcting for the envelope by using trigonometric identities to correct for the phase.

The goal is to recover the potential $V(\boldsymbol{\omega})$. The measured images by tcBF and tcDPC are,
\begin{equation}
    \begin{gathered}
I_{\mathrm{tcBF}}(\boldsymbol{\omega})=V(\boldsymbol{\omega}) \operatorname{PCTF}_{\mathrm{tcBF}}(\boldsymbol{\omega}) ,
\end{gathered}
\end{equation}
\begin{equation}
    \begin{gathered}
I_{\mathrm{tcDPC}}(\boldsymbol{\omega})=V(\boldsymbol{\omega}) \operatorname{PCTF}_{\mathrm{tcDPC}}(\boldsymbol{\omega}).
\end{gathered}
\end{equation}

For simplicity we write $\gamma=-1 / 2 k_0 \Delta f |\boldsymbol{\omega}|^2$ and deconvolve the envelope functions,
\begin{equation}\label{equ:tcbf_image}
    \begin{aligned}
& \frac{I_{\mathrm{tcBF}}(\boldsymbol{\omega})}{\mathbb{A}(\boldsymbol{\omega})}=-V(\boldsymbol{\omega}) \sin \left(1 / 2 k_0 \Delta f |\boldsymbol{\omega}|^2\right)=V(\boldsymbol{\omega}) \sin (\gamma) ,
\end{aligned}
\end{equation}
\begin{equation}\label{equ:tcdpc_image}
    \begin{aligned}
 \frac{I_{\mathrm{tcDPC}}(\boldsymbol{\omega})}{i \mathbb{B}(\boldsymbol{\omega})}=V(\boldsymbol{\omega}) \cos \left(1 / 2 k_0 \Delta f |\boldsymbol{\omega}|^2\right)=V(\boldsymbol{\omega}) \cos (\gamma).
\end{aligned}
\end{equation}

Combining Equation \ref{equ:tcbf_image} and \ref{equ:tcdpc_image} we can extract a complex image with a phase shift determined by the recorded image intensities of $\mathrm{tcBF}, \mathrm{tcDPC}$ and their envelopes respectively:
\begin{equation}
    \frac{I_{t c D P C}(\boldsymbol{\omega})}{i \mathbb{B}(\boldsymbol{\omega})}+i \frac{I_{t c B F}(\boldsymbol{\omega})}{\mathbb{A}(\boldsymbol{\omega})}=V(\boldsymbol{\omega})[\cos (\gamma)+i \sin (\gamma)]=V(\boldsymbol{\omega}) \exp (i \gamma).
\end{equation}

AcBF can be equivalently achieved by correcting for the phase shift $\exp (i \chi)$ in this complex image:
\begin{equation}
    \begin{gathered}
\frac{I_{\mathrm{tcBF}}(\boldsymbol{\omega})}{I_{\mathrm{tcDPC}}(\boldsymbol{\omega})} \frac{i \mathbb{B}(\boldsymbol{\omega})}{\mathbb{A}(\boldsymbol{\omega})}=\tan (\gamma) ,
\end{gathered}
\end{equation}
\begin{equation}
    \begin{gathered}
\exp (i \gamma)=\exp \left(i \tan ^{-1}\left[\frac{I_{\mathrm{t c B F}}(\boldsymbol{\omega})}{I_{\mathrm{t c D P C}}(\boldsymbol{\omega})} \frac{i \mathbb{B}(\boldsymbol{\omega})}{\mathbb{A}(\boldsymbol{\omega})}\right]\right).
\end{gathered}
\end{equation}

And the underlying potential can be recovered by,
\begin{equation}
    V(\boldsymbol{\omega})=\left\{\frac{I_{\mathrm{tcDPC}}(\boldsymbol{\omega})}{\mathbb{B}(\boldsymbol{\omega})}-\frac{I_{\mathrm{tcB} F}(\boldsymbol{\omega})}{\mathbb{A}(\boldsymbol{\omega})}\right\} \exp \left(-i \tan ^{-1}\left[\frac{I_{\mathrm{tcBF}}(\boldsymbol{\omega})}{I_{\mathrm{tcDPC}}(\boldsymbol{\omega})} \frac{i \mathbb{B}(\boldsymbol{\omega})}{\mathbb{A}(\boldsymbol{\omega})}\right]\right) / i.
\end{equation}

In the case of forming a complex image with the recorded tcBF and tcDPC images,
\begin{equation}
\begin{aligned}
I_{\mathrm{t c B F}}(\boldsymbol{\omega})+I_{\mathrm{t c DPC}}(\boldsymbol{\omega})&=\operatorname{PCTF}_{\mathrm{acBF}}(\boldsymbol{\omega}) V(\boldsymbol{\omega})\\
&=\left[\operatorname{PCTF}_{\mathrm{tcBF}}(\boldsymbol{\omega})+\operatorname{PCTF}_{\mathrm{tcDPC}}(\boldsymbol{\omega})\right] V(\boldsymbol{\omega}) ,
\end{aligned}
\end{equation}
\begin{equation}
\begin{aligned}
\operatorname{PCTF}_{\mathrm{acBF}}(\boldsymbol{\omega})=[\mathbb{A}(\boldsymbol{\omega}) \sin (\gamma)+i \mathbb{B}(\boldsymbol{\omega}) \cos (\gamma)],
\end{aligned}
\end{equation}
which after some mathematical manipulation results in the same form as Equation \ref{equ: acbf_ctf}.

\section{Noise considerations in in-focus acBF, DPC, and SSB}\label{appendix_part3_C}
\setcounter{figure}{0} % reset figure numbering
\renewcommand{\thefigure}{C\arabic{figure}} % ensure format is A1, A2, etc.

In the main text, we have assumed a flat Poisson noise power spectrum in the data across all imaging methods discussed. Therefore, the comparison became very simple, by comparing just the squares of the PCTFs. However, different postprocessing of the 4D-STEM data can result in different DQEs, even if the CTF is the same, e.g., in the case of in-focus tcDPC (i.e., DPC) v.s. SSB. Bennemann et al \cite{bennemann_detective_2025} derives the DQE for in-focus and defocused SSB. Here we discuss in more details how the dose efficiency can be further improved by precisely constraining the phase contrast transferring regions.

Within in the bright field disk, the phase information is encoded in the triple overlap (TO) region and two sidebands (double overlap (DO) regions). The two DO regions contain repeated information, thus SSB requires analysis of only one of the sidebands. When both TO and DO regions are used, e.g., acBF, the NPS is given by,
\begin{equation}
    \left|\operatorname{N P S}_{\mathrm{acBF}}(\boldsymbol{\omega})\right|^2=\mathbb{A}(\boldsymbol{\omega})+\mathbb{B}(\boldsymbol{\omega}).
\end{equation}

That the NPS equals the sum of the CTF envelopes is a result from the mean and variance being equal in the Poisson model.

Following Section \ref{sec: wpoa_contrast}, the phase contrast transfer in DPC and SSB is the same, because both are utilizing only the double overlap regions. However, in contrary to a plain subtraction of the two halves of the disks as in DPC, SSB precisely cuts out the double overlap trotters. The effect of this processing results in different NPS's.
\begin{equation}
    \begin{gathered}
\left|\operatorname{N P S}_{\mathrm{DPC}}(\boldsymbol{\omega})\right|^2=\mathbb{A}(\boldsymbol{\omega}) / 2+\mathbb{B}(\boldsymbol{\omega}) ,
\end{gathered}
\end{equation}
\begin{equation}
    \begin{gathered}
\left|\operatorname{N P S}_{\mathrm{SSB}}(\boldsymbol{\omega})\right|^2=\mathbb{B}(\boldsymbol{\omega}).
\end{gathered}
\end{equation}

\begin{figure}[h]
    \centering
    \includegraphics[width=0.7\linewidth]{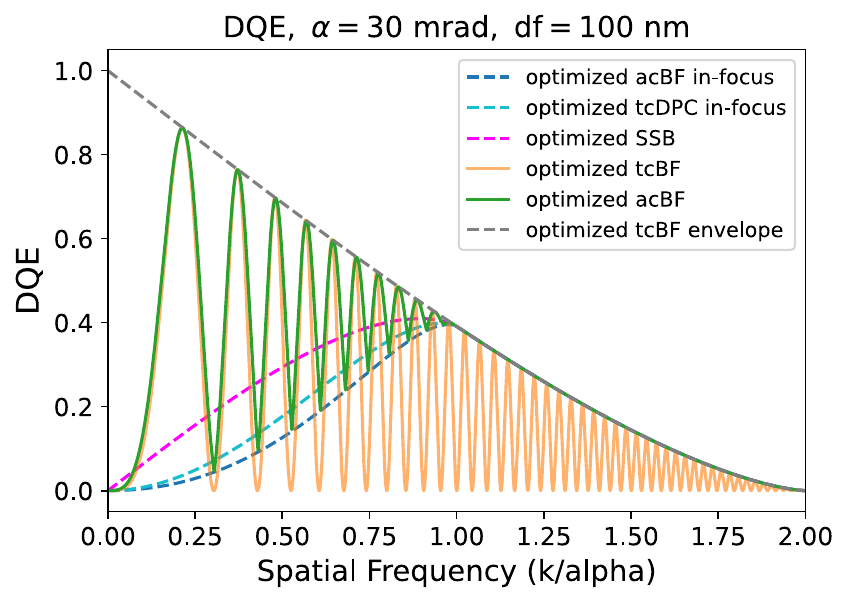}
    \caption{DQE of in-focus tcDPC (DPC), SSB, tcBF and acBF. Precisely cutting the DO regions can slightly improve dose efficiency.}
    \label{fig:info_C1}
\end{figure}

Figure \ref{fig:info_C1} shows the resultant DQEs of all the phase retrieval methods discussed in this paper under this Poisson noise assumption. Note that this treatment should only be viewed qualitatively, as the single overlap region (white area in Figure \ref{fig:info_fig2}) has been neglected, although it is a small portion in the BF disk and simply the transmitted beam, thus not producing phase contrast transfer. Bennemann et al \cite{bennemann_detective_2025} carries out similar analysis of this noise consideration.

%%       \cite{<label>} ==> [1]

%%
% Example citation, See \cite{lamport94}.





% %% For numbered reference style

%   \textit{\LaTeX: a document preparation system},
%   Addison Wesley, Massachusetts,

% \end{thebibliography}
% \bibliographystyle{unsrt}
\bibliographystyle{unsrt}
\bibliography{info_4d}

\end{document}